\documentclass[aps,pre,amsmath,amsfonts,floatfix,superscriptaddress]{revtex4}
\usepackage{graphicx}

\let\a=\alpha   \let\g=\gamma  \let\d=\delta \let\e=\varepsilon
\let\z=\zeta     \let\l=\lambda
    \let\n=\nu         \let\p=\pi   
    \let\f=\varphi 
   
\let\G=\Gamma \let\D=\Delta   
    \let\Si=\Sigma     
  \let\r=\rho \let\th=\theta
\let\io=\infty
\def\ie{{i.e. }}\def\eg{{e.g. }}

\def\PP{{\cal P}}

\def\ZZ{{\cal Z}}

\def\to{\rightarrow}
\def\la{\left\langle}
\def\ra{\right\rangle}

\newcommand{\beq}{\begin{equation}}
\newcommand{\eeq}{\end{equation}}

\begin{document}

\title{On the solution of a `solvable' model of an ideal glass of hard spheres \\ displaying a jamming transition}

\date{\today}

\author{Marc M\'ezard}
\affiliation{
Laboratoire de Physique Th\'eorique et Mod\`eles Statistiques,
Batiment 100, Univ. Paris Sud and CNRS,
F-91405 Orsay, France 
}

\author{Giorgio Parisi}
\affiliation{Dipartimento di Fisica, 
Sapienza Universit\'a di Roma,
INFN, Sezione di Roma I, IPFC -- CNR,
P.le A. Moro 2, I-00185 Roma, Italy
}

\author{Marco Tarzia}
\affiliation{Laboratoire de Physique Th\'eorique de la Mati\`ere Condens\'ee, Universit\'e Pierre et Marie Curie-Paris 6,
UMR CNRS 7600, 4 place Jussieu, F-75252 Paris Cedex 05, France.}

\author{Francesco Zamponi}
\affiliation{Laboratoire de Physique Th\'eorique, 
\'Ecole Normale Sup\'erieure, UMR CNRS 8549,
24 Rue Lhomond, F-75231 Paris Cedex 05, France
}

\begin{abstract}
We discuss the analytical solution through the cavity method of a mean field
model that displays at the same time an ideal glass transition and a set of
jamming points. We establish  the equations describing this system,
and we discuss some approximate analytical solutions
and a numerical strategy to solve them exactly. We compare these methods and we get
insight into the reliability of the theory for the description of finite dimensional hard spheres.
\end{abstract}

\maketitle

\tableofcontents

\section{Introduction}

The theoretical investigation of the glass transition and its relation to jamming
in hard sphere systems has made considerable progress in the last 30 
years~\cite{SW84,SSW85,Sp98,CFP98,PZ10}. This has been possible mainly
because of the powerful analogy between jammed states and inherent 
structures~\cite{SW82,LS90,Sp98,KK07} and of the development of methods based
on spin glass theory~\cite{Mo95,MP99} to describe the glass transition
of particle systems. This progress led to the proposal that amorphous
jammed states of hard spheres can be thought of as the states obtained
in the infinite pressure limit
of metastable glasses, and therefore described using tools of (metastable-)equilibrium
statistical mechanics.

\begin{figure}
\includegraphics[width=8cm]{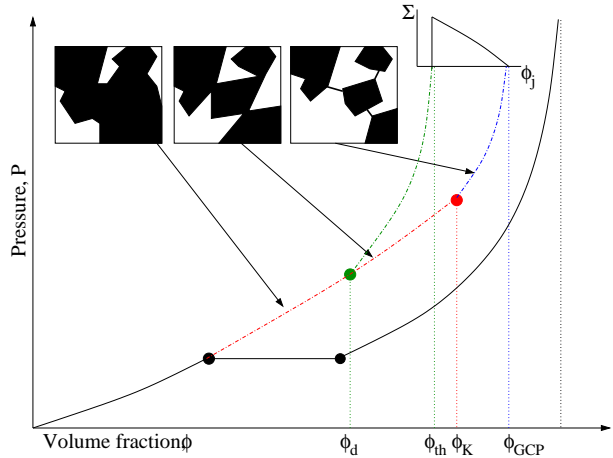}
\caption{
Schematic mean-field phase diagram of hard spheres in three dimensions,
see the text and~\cite{PZ10} for a detailed description. 
}
\label{dia_totale}
\end{figure}

The phase diagram of hard spheres 
that results from these mean-field studies is summarized in Fig.~\ref{dia_totale},
where we plot the pressure as a function of the packing fraction $\f$
which is the fraction of space covered by the spheres. 
The full black line represents the
equilibrium phase diagram with the liquid-to-crystal transition.
If this transition can be avoided (by compressing fast enough or by introducing
some degree of polydispersity), one enters into a metastable liquid
phase. The nature of this metastable liquid changes at $\f=\f_{\rm d}$.
It consists of a single ergodic state for $\f<\f_{\rm d}$. When $\f>\f_{\rm
  d}$, the available phase space splits into many glassy states. If the
system is stuck in one of these states and compressed, it follows one of
the glass branches of the phase diagram, until its pressure eventually diverges at
some packing fraction $\f_j$ which depends on the state. 
At density $\f_{\rm K}$ a thermodynamic glass transition happens
(in the sense of mean field spin glasses~\cite{CC05}) towards an {\it ideal glass}.
The pressure of the latter diverges at $\f_{\rm GCP}$. In the inset, the complexity,
\ie the logarithm of the number of glassy states, is plotted as function of
the jamming density $\f_j$: this approach predicts that there exist jammed states in
a finite interval of density $\f_j \in [\f_{\rm th}, \f_{\rm K}]$.
The boxes show a schematic picture of the ($3N$-dimensional,
where $N$ is the number of particles) 
phase space of the system: black configurations are allowed by the
hard-core constraint, white ones are forbidden. In the supercooled liquid phase the allowed configurations form a connected
domain; however, on approaching $\f_{\rm d}$ the connections between different metastable regions 
become smaller and smaller. Above $\f_{\rm K}$, they disappear in the thermodynamic limit and glassy
states are well defined.

The above mean-field picture has been obtained by a succession of
works which start from the studies of some categories of spin-glasses
with so-called `one step replica symmetry breaking', and have
gradually matured into analytic approximation tools for the theory of
hard spheres (see \cite{PZ10} and references therein).
A very interesting model has been introduced recently by Mari,
Krzakala and Kurchan~\cite{MKK08}.
It displays exactly the phase diagram presented in Fig.~\ref{dia_totale}:
it undergoes an equilibrium 
glass transition and it has an interval of densities where it shows
all the phenomenology which is now associated to jamming, like
marginal mechanical stability and the associated presence of anomalous soft
modes in the vibrational spectrum~\cite{OSLN03,Wyart,WNW05}. 
The model has been studied
numerically in~\cite{MKK08} in order to show the existence of separate
glass and jamming transitions and to clarify to some extent the relation between
the two.

This model is interesting in that it is in principle solvable: it can
be investigated by mean of modern methods that have been developed in the context
of mean field spin glasses, the replica method~\cite{MPV87} and the
cavity method~\cite{MM09}. This investigation  is the purpose of the present paper, where 
we derive the cavity equations that describe the model and we present some
approximated analytical 
solutions to them, along with a detailed numerical resolution.
Since it will turn out that the exact solution requires quite heavy numerical
calculations (heavier than a direct Monte Carlo study of the model, at least
for a moderate number of particles,
such as the one performed in~\cite{MKK08}), one might wonder why this 
solution is interesting at all. There are at least two reasons why this study
is interesting, in our opinion.
The first is that Monte Carlo methods are not able to access the deep
glassy phase or the densest part of the jammed phase: they are confined to explore the region close to $\f_d$
(at equilibrium) and $\f_{\rm th}$ (at jamming). Therefore if one wants
to study, for instance, how the properties of the packings change when going
from $\f_{\rm th}$ to $\f_{\rm GCP}$, the exact solution is needed.
Moreover, we will show that the cavity method allows to derive simple analytical
approximations to the true solution. Similar approximations have been used to study finite dimensional hard spheres~\cite{PZ10};
their investigation in the controlled setting of the present
'solvable' model allows to assess their reliability.
Finally, there are some generic structures in the correlations of jammed packings
that one would like to explain analytically. Our work is a first step
in this direction.

This paper is meant to be read by specialists in the field, so we did not make
much attempt to explain in details the basis of the method. Recent complete reviews of
the physical problem~\cite{PZ10,He10,LNSW10,TS10} as well as of the method we 
used~\cite{Pa07b,MM09}
exist, and the reader is assumed to be familiar with these concepts.

\section{Definitions}

\begin{figure}
\includegraphics[width=5cm,angle=90]{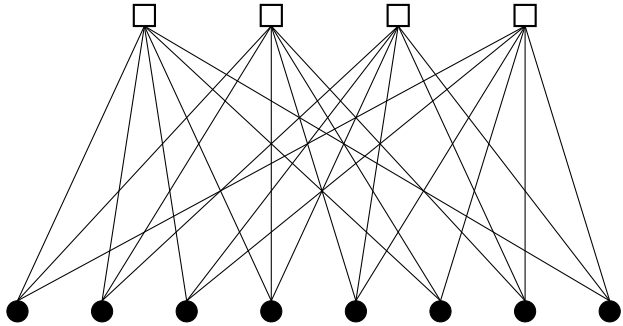}
\caption{
An illustration of the model for $p=6$, $z=3$ and $N=8$. Each white
square is a box, each black dot is a variable (sphere).
Each box contains all the spheres connected
to it by a link. The sphere inside one box must not overlap
(note that for $z=1$ one obtains $N/p$ systems of $p$ hard spheres).
}
\label{factor_graph}
\end{figure}

The model that we study in this paper is a simple generalization of the one introduced
in~\cite{MKK08},  defined as follows.
We consider a ``factor graph'', namely a bipartite graph
made by two types of nodes: {\it variables} and {\it boxes}.
Each variable is connected to $z$ boxes and each box is 
connected to $p$ variables.
In a system with $N$ variables the number of boxes is $Nz/p$ and the total
number of links (\ie variable-box connections) is $Nz$. We will consider
an ensemble of `random regular' factor graphs where each graph satisfying this requirement
has the same probability. A crucial properties of this ensemble, that allows
for the solution of the model, is that in the thermodynamic limit $N\to\io$
almost all graphs are locally tree-like, in a sense that can be defined 
precisely~\cite{MM09}.

Each variable is a vector $x_i \in [0,1]^d$ with periodic boundary conditions, 
where $d$ is the {\it dimension} and $i = 1, \cdots, N$.
In the following we denote by 
$|x_i - x_j| = \sqrt{\sum_{\mu=1}^d( |x_i^\mu - x_j^\mu|_{\text{mod }1})^2}$ 
the distance between $x_i$ and its closest periodic image of $x_j$.
If we call $\chi(x_i,x_j)$ the characteristic function of the hard sphere constraint
(with periodic boundary conditions), 
\ie $\chi(x_i,x_j)=1$ if $|x_i - x_j| \geq D$ and $0$ otherwise,
then each box $a = 1, \cdots, Nz/p$ imposes the condition
\beq
\chi(a) \equiv \chi(x^a_1,\cdots,x^a_p) \equiv \prod_{i<j}^{1,p} \chi(x^a_i,x^a_j) \neq 0 \ ,
\eeq
where $x^a_i$ are the variables connected to box $a$.
The partition function of the model is
\beq
Z = \int dx_1 \cdots dx_N \prod_{a=1}^{Nz/p} \chi(a) \ .
\eeq

A pictorial description of the model is the following (see Fig.~\ref{factor_graph}). 
Each box can be thought
of as a cubic region $[0,1]^d$ with periodic boundary conditions. 
Each variable node $i=1,\dots,N$ represents a ``sphere'' of diameter
$D$ and this sphere appears in position $x_i$ in
all the $z$ boxes to which the node is connected. On the other
hand, each box contains exactly $p$ spheres. The constraint is that, for each
box, the $p$ spheres present in the box do not overlap.

The model therefore differs from a standard hard sphere model, since each
sphere interacts only with a finite subset of neighbors, and the topology of 
the interaction network is fixed by the random graph construction described 
above. This structure is such that the model becomes a mean field model
and is therefore exactly solvable, at least in principle, as we will discuss in the
following. It is worth to note, however, that there are two ``formal'' limits where
one gets back the standard hard sphere model:
in the case $z=1$ the model reduces to $N/p$ independent
systems of $p$ hard spheres each, while for $p=2$ and $z=N-1$ one
gets back a single system of $N$ hard spheres. Note also that in \cite{MKK08}
only the version with $p=2$ has been studied.

Our investigations showed, however, that the model defined above undergoes
a ``crystallization'' phenomenon at high density: the spheres tend to localize
around a discrete set of positions inside the unit box. This has been avoided
in~\cite{MKK08} by introducing a small degree of polydispersity of the size of
spheres. Here, in the analytical treatment of the model, we do not need to
use this trick since we can impose directly that the solutions are translationally
invariant, therefore discarding all crystalline phase of the model. In this way
one effectively restricts to the amorphous phases, but one should keep in mind
that these are metastable with respect to the crystal in the true model.
Another possibility to remove the non-translationally invariant phase is to 
introduce local ``random shifts'': on each link we introduce a quenched variable
$s_{ai} \in [0,1]^d$,
such that the corresponding particle appears in the corresponding box translated
by $s_{ai}$. On a tree with open boundary conditions, 
this will not change the model since one can always perform a change of variable
to remove the shifts. In presence of loops however, the random shifts will frustrate
the periodic order. But since the cavity solution is based
on local recursions, the solutions describing the model with random shifts will be
the same as the translationally invariant solutions of the model without random
shifts.
A similar situation occurs when studying an antiferromagnetic model on
a random graph: local recursion relations allow both an
antiferromagnetic and an amorphous ordering. The former is irrelevant
on a random graph because long loops of odd length frustrate the antiferromagnetic
order. The antiferromagnetic system thus behaves like the spin glass in
which the sign of the couplings are quenched random variables.
See Ref.~\cite{col2} for a more detailed discussion in the context of a very similar
model.

We define $V_d(R)$ the volume of a $d$-dimensional hypersphere of radius $R$; then
$V_s = V_d(D/2) = 2^{-d} V_d(D)$ is the volume of one hard sphere (since the spheres
have diameter $D$),
and $\f = p V_s$ is the packing fraction, that represents the fraction of the unit
box that is covered by the $p$ interacting spheres. It is trivial to check that there
are no configurations with $\f > 1$.
The parameter that controls
the packing fraction is the diameter $D$ since the box size is fixed; for this reason
in the following we will use directly the sphere diameter $D$ as control parameter and label the different
transitions as $D_{\rm K}$, $D_{\rm GCP}$, $D_{\rm d}$, etc.

For a system of $p$ hard spheres in $d$ dimensions, we define the following quantities:
\beq\begin{split}
&Z^0_p =  \int dx_{1}\cdots dx_{p} \prod_{i<j}^{1,p} \chi(x_{i},x_{j}) \ , \\
&g^0_p(x-y) = \frac1{p(p-1)} \la \sum_{i\neq j}^{1,p} \d(x-x_i)\d(y-x_j) \ra 
= \frac{1}{Z^0_p} \int dx_3 \cdots dx_p \chi(x,y,x_3,\cdots,x_p)
\ , 
\end{split}\eeq
such that $Z^0_p$ is the partition function of $p$ hard spheres (apart from a $p!$),
and $g^0_p$ is related to the usual pair correlation function~\cite{Hansen} by
\beq
g(r) = \frac{p-1}{p} g^0_p(r) \ .
\eeq
For the following discussion, it will be useful to define
\beq\label{voidspace}
v_n(x_1,\cdots,x_n) = \int dx \prod_{i=1}^n \chi(x,x_i)
\eeq
which is the so called {\it void space} or {\it cavity volume}, namely
the volume available to insert an additional sphere in a box 
given the positions
of $n$ other spheres, $\{x_1, \cdots, x_n \}$.

\section{Cavity equations}
\label{sec:cavityeq}

The cavity method has now become a standard method to solve statistical models
defined on random graphs. 
We will not explain here the method and refer the reader to~\cite{MM09,cavity}. 
Here we only
write the equations for our specific case.

\subsection{Bethe free energy}

We define by $\partial i$ the set of boxes connected to variable $i$, and by
$\partial a$ the set of variables connected to box $a$.
On each link we define two fields: $\f_{a\to i}(x_i)$ is the
probability density of the variable
$x_i$ when connected only to the box $a$; $\psi_{i\to a}(x_i)$ is the
probability density
of the same variable when connected to all the boxes in its neighborhood but $a$.
Both are normalized to 1 and
they satisfy the equations:
\beq\label{recurrence}
\begin{split}
&\psi_{i\to a}(x_i) = \frac{1}{Z_{i\to a}} \prod_{b \in \partial i \setminus a} \f_{b\to i}(x_i) \ , \\
&\f_{a\to i}(x_i) = \frac{1}{Z_{a\to i}} \int 
\left(\prod_{j\in \partial a \setminus i} dx_j \psi_{j\to a}(x_j) \right) \chi(a) \ ,
\end{split}
\eeq
which can derived from the stationarity of the Bethe entropy:
\beq\label{Sbethegraph}
S = - \sum_{\text{links }a-i} \log \int dx_i \psi_{i\to a}(x_i) \f_{a\to i}(x_i)
+ \sum_a \log \int 
\left(\prod_{j\in \partial a} dx_j \psi_{j\to a}(x_j) \right) \chi(a)
+\sum_i \log \int dx_i \prod_{a \in \partial i } \f_{a\to i}(x_i) \ .
\eeq
These equations have the general form of the cavity (or Bethe) equations that can be derived
for any model with local interactions~\cite{MM09}. With respect to
previous studies of frustrated systems with the cavity method, the main difference here (and the main
source of difficulty) is the fact that the variables $x$ are {\it continuous}.
Although the Bethe free energy is not variational in general, it has the property that the
cavity equations can be obtained imposing its stationarity with respect to the cavity fields.
In some special cases one can argue that it provides indeed an upper or lower bound to the
true free energy, but a proof of this is still lacking.

\subsection{Replica symmetric cavity equations}

The replica symmetric (RS) equations for such a regular graph are trivially obtained by dropping
the spatial dependence of the fields. In this case we use the notation
$Z_\f = Z_{a\to i}$ and $Z_\psi = Z_{i\to a}$, and we get
\beq\label{RSitera}
\begin{split}
&\psi(x) = \frac{1}{Z^{RS}_{\psi}} \f(x)^{z-1} \ , \\
&\f(x) = \frac{1}{Z^{RS}_{\f}} \int 
\left(\prod_{j=1}^{p-1} dx_j \psi(x_j) \right) \chi(x,x_1,\cdots,x_{p-1}) \ ,
\end{split}
\eeq
and the RS entropy per particle is
\beq
S_{RS} = - z \log \int dx \psi(x) \f(x)
+ \frac{z}p \log \int 
\left(\prod_{j=1}^p dx_j \psi(x_j) \right) \chi(x_1,\cdots,x_p)
+ \log \int dx \f(x)^z \ .
\eeq

These equations admit the trivial translationally invariant solution
$\psi(x)=\f(x)=1$ with $Z^{RS}_\psi = 1$ and 
\beq
Z^{RS}_\f = \int dx \int \left( \prod_{j=1}^{p-1} dx_j \psi(x_j)
  \right) \chi(x,x_1,\cdots,x_{p-1}) \equiv Z^0_p \ ,
\eeq
that is the partition function of $p$ Hard Spheres in the unit box.
Therefore the entropy of the RS phase is
\beq\label{SRS}
S_{RS} = \frac{z}{p} \log  Z^0_p \ .
\eeq

\subsection{1-Step replica symmetry breaking cavity equations}

In the standard interpretation~\cite{MM09}, the glass phase is signaled
by the appearance of multiple solutions $\psi_{i\to a}^{(\a)}$,
$\f_{a\to i}^{(\a)}$, of Eq.~(\ref{recurrence}).
Each of these solutions represents a glass state with entropy $s_\a$ given by the
Bethe entropy (\ref{Sbethegraph}) computed on the corresponding set of fields.
Although one does not have direct access to individual glassy solutions (since the
direct numerical solution of the Bethe equations by iteration on a single graph is extremely unstable
in this region), a statistical treatment of the properties of the solutions in this regime
exists and goes under the name of 1-step replica symmetry breaking (1RSB) 
description~\cite{cavity}.
It is based on an
entropy $S(m)$ which is the sum over all solutions $\a$
of the corresponding partition 
function $Z_\a = e^{N s_\a}$ to power $m$~\cite{Mo95}. 
The latter is computed by looking to the evolution of the solutions of the Bethe equations
under an iteration that adds one more variable to the graph~\cite{cavity}, or more simply by
introducing an auxiliary model and assuming that a RS description holds for that model~\cite{MM09}. 
We do not discuss here these derivations and only report the resulting equations for our model,
which are the following:
\beq\label{S1RSB}
\begin{split}
& S(m) = \frac{1}N \log \sum_\a Z_\a^m = m s(m) + \Si(m) = -z S_{link}(m) + \frac{z}p S_{box}(m) + S_{site}(m) \\
& S_{link}(m) = \log \int d\PP[\psi] d\PP[\f] \left[ \int dx \psi(x) \f(x)  \right]^m
\equiv \log \la Z_{link}^m \ra \ , \\
& S_{box}(m) = \log \int d\PP[\psi_1]\cdots d\PP[\psi_p] 
\left[ \int \left( \prod_{j=1}^p \psi_{j}(x_j) dx_j \right) \chi(x_1,\cdots,x_p) \right]^m 
\equiv \log \la Z_{box}^m \ra \ , \\
& S_{site}(m) = \log \int d\PP[\f_1]\cdots d\PP[\f_z] 
\left[ \int dx \prod_{i=1}^z \f_i(x) \right]^m \equiv \log \la Z_{site}^m \ra\ .
\end{split}\eeq
The stationarity of this function with respect to $\PP[\psi]$ and $\PP[\f]$ gives the 1RSB equations:
\beq\label{rec1RSB}
\begin{split}
& \PP[\psi] = \frac1{\ZZ_\psi} \int \prod_{i=1}^{z-1} d\PP[\f_i] 
\d\left[ \psi(x) - \frac{1}{Z_\psi} \prod_{i} \f_{i}(x) \right] (Z_\psi)^m \ , \\
& \PP[\f] = \frac1{\ZZ_\f} \int \prod_{i=1}^{p-1} d\PP[\psi_i]
 \d\left[ \f(x) - \frac{1}{Z_\f} \int 
\prod_j dx_j \psi_j(x_j) \chi(x,x_1,\cdots,x_{p-1}) \right] (Z_\f)^m \ .
\end{split}\eeq
where the normalization constants are
\beq\label{reweighting}
\begin{split}
& Z_{\psi}[\f_1,\cdots,\f_{z-1}] = \int dx \prod_{i} \f_{i}(x) \ , \\
& Z_\f[\psi_1,\cdots,\psi_{p-1}] = \int dx \prod_j dx_j \psi_j(x_j) \chi(x,x_1,\cdots,x_{p-1}) \ , \\
& \ZZ_\psi = \la (Z_\psi)^m \ra \ , \\
& \ZZ_\f = \la (Z_\f)^m \ra \ .
\end{split}\eeq
The internal entropy can then be written, using the standard method of~\cite{Mo95}, as
\beq\label{sSig1RSB}
\begin{split}
& s(m) = \frac{\partial S(m)}{\partial m} = 
-z \frac{\la Z_{link}^m \log Z_{link}\ra }{\la Z_{link}^m \ra}
+ \frac{z}p \frac{\la Z_{box}^m  \log Z_{box}\ra }{\la Z_{box}^m \ra} 
+ \frac{\la Z_{site}^m \log Z_{site}\ra }{\la Z_{site}^m \ra} \\
\end{split}\eeq
and the complexity is $\Si(m) = S(m) - m s(m)$.
The parameter $m$ is the 1RSB parameter, whose equilibrium value must be
fixed imposing that the replicated entropy is stationary~\cite{MPV87}.


\section{The stability of the RS solution}
\label{sec:RSstability}

To study the stability of the RS phase we perturb around it:
\beq
\psi_{i\to a}(x) = 1 + A e^{-ikx + i\th_{i\to a}} \ ,
\eeq
and look at the linear stability of $A$ assuming that the phase $\th$ is
random, \ie when substituting in the right hand side of (\ref{RSitera})
each $\psi$ get a random independent phase. This is done in order to enforce
translational invariance, otherwise we would study the instability towards
modulated phases, which is indeed interesting but we do not consider here, for
reasons discussed in the introduction.
Note that we have $k = 2 \pi (n_1,\cdots,n_d)$, where $n_i$ are integer numbers.
Then at first order we have
\beq\label{STlin}
A e^{-i k x + i \th} = A \frac{1}{Z^0_p} \sum_{a=1}^{z-1} \sum_{j=1}^{p-1}
\int dx_2 \cdots dx_{p} \chi(x,x_2,\cdots,x_{p}) e^{-i k x_2 + i \th_{j \to a}} \ .
\eeq
Now we can bring the factor $e^{-ikx}$ on the other side and integrate over
$x$; moreover we take the square and use that the $\th_{j \to a}$ are random and
uncorrelated and we obtain the final result
\beq
A^2 = A^2 (z-1)(p-1) \left| \frac{1}{Z^0_p}\int dx_1 \cdots dx_{p}
  \chi(x_1,\cdots,x_{p-1}) e^{i k (x_1-x_2)} \right|^2 \ .
\eeq
Defining
\beq\begin{split}
&g^0_p(k) = \int dx dy e^{ik(x-y)} g^0_p(x-y) =
\frac{1}{Z^0_p}\int dx_1 \cdots dx_{p}
  \chi(x_1,\cdots,x_{p}) e^{i k (x_1-x_2)} \ ,
\end{split}\eeq
the stability condition is
\beq\label{RSstabcondition}
\sqrt{(p-1)(z-1)} |g^0_p(k)| \leq 1 \ , \hskip2cm \forall k =2 \pi (n_1,\cdots,n_d) \neq 0 \ .
\eeq
Hence from the knowledge of $Z^0_p$ and $g^0_p(k)$ 
we can compute the RS entropy and the stability of the RS solution.

\subsection{Results for $p=2$, any dimension}

For $p =2$, $k\neq 0$ and $D < 1/2$, we have simply $g^0_2(x-y) = \chi(x-y)/(1-V_d(D))$ and
\beq
g^0_2(k) = \int_{[0,1]^d} dx \, \frac{e^{i k x} \chi(x)}{1-V_d(D)} = 
-\int_{[-1/2,1/2]^d} dx \, \frac{e^{i k x} \th(|x|<D)}{1-V_d(D)} = 
- \left(\frac{2\pi D}k\right)^{d/2} \frac{J_{d/2}(k D)}{1-V_d(D)}
\ .
\eeq
One can show that for the values of $D$ we are interested in, the maximum
of $g^0_2(k)$ is assumed for $k = 2 \pi$, \ie the smallest $k$.
Then the condition on $D$ is
\beq
\frac{D^{d/2} J_{d/2}(2\pi D)}{1-V_d(D)} \leq \frac1{\sqrt{z-1}} \ .
\eeq
In the limit $z\to\io$, as $D$ is small, 
we can use $J_{n}(x) \sim (x/2)^{n} / \G(n+1)$, and
neglecting the denominator
\beq
\frac{D^{d/2} J_{d/2}(2\pi D)}{1-V_d(D)} \sim \frac{\p^{d/2} D^d}{\G(d/2+1)}
= V_d(D) \leq \frac1{\sqrt{z-1}} \ .
\eeq

\subsection{Results for $d=1$, any $p$}

In $d=1$ we get, from the exact solution
\beq\begin{split}
& Z^0_p = [1-p D]^{p-1}  \ , \\
& g^0_p(k) = \frac{1}{p-1} \sum_{n=0}^{p-2} e^{-i (n+1) k D} 
\ _1 F_1[1+n;p;-i(1-p D) k]    \ .
\end{split}\eeq
where $_1 F_1[a;b;z]$ is the confluent hypergeometric function of the first
kind. Also in this case the lowest $k$ becomes unstable in the first place.

\subsection{Results for $d=2$ and $p=3$}

As a last interesting case, we consider $d=2$ and $p=3$.
In the following for simplicity 
we consider $D<1/4$ to avoid problems coming from periodic boundary
conditions.

We start by the computation of the partition function $Z^0_3$
of three spheres in a box, which can be done using the standard virial expansion.
For convenience we fix the first sphere,
as well as the origin of the coordinate frame, in the center of the box.
The center of the second sphere can be anywhere in the box outside a disk of
radius $D$ centered in the origin. 
Given the position of the second sphere,
the third sphere can be anywhere outside the union of two disks centered
around the first two spheres. 

If the second sphere is at distance $r = |x_2-x_1|$ from the origin $x_1=0$, the free volume accessible
to the third sphere is
\beq
v_2(x_1,x_2) = 1 - 2 \pi D^2 + \theta(2D-r) D^2 \left( 2 \arccos \frac{r}{2D} - \frac{r}{2D} \sqrt{4 - \frac{r^2}{D^2}} \right)
\eeq
This has to be integrated over the position of the second sphere. There are three possible cases:
\begin{enumerate}
\item $r \in [D,2D]$; in this case the first and second exclusion spheres have an overlap, and the second sphere
can rotate at any angle without hitting the boundary of the box. Therefore one has
\beq
Z^0_3(1) = 2\pi \int_D^{2D} dr \, r \left[1 - 2 \pi D^2 +D^2 \left( 2 \arccos \frac{r}{2D} - \frac{r}{2D} \sqrt{4 - \frac{r^2}{D^2}} \right) \right]
\eeq
\item $r \in [2 D, 1/2]$ (recall that the box has side 1 so $r$ is at most $1/2$); in this case the first and second exclusion spheres have no overlap, and the second sphere can rotate
at any angle, therefore
\beq
Z^0_3(2) = 2\pi \int_{2D}^{1/2} dr \, r \left(1 - 2 \pi D^2 \right)
\eeq
\item $r \in [1/2,\sqrt{2}/2]$; also in this case there is no overlap contribution, but the second sphere can only be at some angles because of the cubic shape
of the box. 
The total angle that can be spanned is $8 (\pi/4 - \arccos(1/(2r)))$, therefore
\beq
Z^0_3(3) = 8 \int_{1/2}^{\sqrt{2}/2} dr \, r \left(1 - 2 \pi D^2 \right)  \left(\frac{\pi}4 - \arccos  \left(  \frac1{2r}  \right)  \right)
\eeq
\end{enumerate}
All the integrals can be evaluated and summing the three contributions one gets the final result
\beq\label{Z03}
Z^0_3 = 1 - 3 \pi D^2 + \frac14 \pi D^4 \big( 3 \sqrt{3} + 8 \pi \big) \ ,
\hskip2cm
D < 1/4 \ .
\eeq

We also need the value of the pair correlation at contact, $g^0_3(D)$. Following the same reasoning this is given by
\beq
g^0_3(D) = \frac{v_2(r=D)}{Z^0_3} = \frac{1 - 2 \pi D^2 + D^2 \left( \frac{2 \pi}{3} - \frac{\sqrt{3}}{2}  \right)}{1 - 3 \pi D^2 + \frac14 \pi D^4 \big( 3 \sqrt{3} + 8 \pi \big)} \ ,
\hskip2cm
D < 1/4 \ .
\eeq
Finally, $g^0_3(x-y) = v_2(x,y)/Z^0_3$, from which one can compute $g^0_3(k)$ numerically and determine
the stability of the RS solution.


\section{The Gaussian approximation}

We now introduce an approximation to describe the 1RSB phase of the model. We assume that
the fields $\psi_j(x)$ and $\f_i(x)$ are 
localized around a position which is randomly distributed
in the box (this maintains the global translational invariance).
This Ansatz, of course, is not a solution of the 1RSB equations. 
However, we expect that it provides a reasonable estimate of $S(m)$, 
which is expected to become more and
more accurate for large connectivity and close to the random close-packing
point. Moreover, we will see in the following, that even if
the variational nature of the replicated entropy cannot be proven,
these approximations give upper bounds for $D_{\rm K}$.
For this reason we will refer from
now on to these approximations as ``variational'' approximations.
Note that if a variational approximation predicts that the Kauzmann
radius is less than the radius where the RS solution is unstable, $D_{\rm K} < D_{\rm RS}$,
then we know for sure that there is a discontinuous transition
occuring at a value of $D$ smaller than $D_{\rm RS} $.

We assume a Gaussian shape for the fields, which leads
to the following assumption for their distribution:
\beq\begin{split}
&\PP[\psi] = \int dX \, \d\left[ \psi(x) - \frac{e^{-\frac{(x-X)^2}{2 A}}}{(2 \p
    A)^{d/2}}  \right] \ , \\
&\PP[\f] = \int dX \, \d\left[ \f(x) - \frac{e^{-\frac{(x-X)^2}{2 \d A}}}{(2 \p
   \d A)^{d/2}}  \right] \ .
\end{split}\eeq
We substitute this Ansatz in the Bethe free energy (\ref{S1RSB}) and determine
the variational parameters $A$ and $\d$ by its extremization.
In the following we will use the definition 
$\g_A(x) = \frac{e^{-\frac{x^2}{2 A}}}{(2 \p A)^{d/2}}$.
Substituting the expressions above in (\ref{S1RSB}), we obtain the following
results:
\beq\begin{split}
&S_{link} = \log \left[ m^{-d/2} [2 \pi (1+\d) A]^{d (1-m)/2}  \right] \ , \\
&S_{site} = \log \left[ m^{(1-z)d/2} z^{(1-m)d/2} (2\pi \d A)^{-(1-m)(1-z)d/2}
\right] \ . \\
\end{split}
\eeq
Note that $S_{box}$ does not depend on $\d$. Therefore we first write the
contribution of $S_{link}$ and $S_{site}$ and optimize with respect to $\d$:
\beq
S_{site}-z S_{link} = -\frac{d}2 (1-m) \log (2 \pi A)
+\frac{d}2 \log m + \frac{d}2(1-m) \log \left[ \frac{z \d^{z-1}}{(1+\d)^z}
\right] \ . 
\eeq
The optimization is straightforward and gives $\d = z-1$ as expected from
the first Eq.~(\ref{recurrence}). The optimized result is
\beq
S_{site}-z S_{link} = -\frac{d}2 (1-m) \log (2 \pi A)
+\frac{d}2 \log m + \frac{d}2(1-m) (z-1) \log \left[ 1-\frac{1}z
\right] \ . 
\eeq
The last term to be computed is $S_{box}$, which has the form:
\beq
S_{box} = \log \int dX_1 \cdots dX_p \left[ \int dx_1\cdots dx_p
  \g_A(x_1-X_1) \cdots \g_A(x_p-X_p) \chi(x_1,\cdots,x_p) \right]^m
\eeq
Unfortunately this cannot be computed exactly and we have to resort to
further approximations.

\subsection{Small cage expansion, first order}

The small cage expansion proceeds as follows~\cite{PZ10}. First we assume that $m$ is
an integer and write $S_{box}$ as:
\beq
S_{box} = \log \int d\bar x_1 \cdots d\bar x_p \r(\bar x_1) \cdots \r(\bar
x_p) \prod_{i<j}^{1,p} \bar \chi(\bar x_i,\bar x_j) \ ,
\eeq
where $\bar x = (x_1,\cdots,x_m)$ is the coordinate of a ``molecule'' made of 
$m$ particles,
$\bar \chi(\bar x,\bar y) = \prod_{a=1}^m \chi(x_a,y_a)$, and
$\r(\bar x) = \int dX \prod_{a=1}^m \g_A(x_a - X)$.
Observing that $\int dx_2 \cdots dx_m \r(\bar x) = 1$,
we write
\beq\begin{split}
S_{box} &= \log \int d\bar x_1 \cdots d\bar x_p \r(\bar x_1) \cdots \r(\bar
x_p) \prod_{i<j}^{1,p} [\bar \chi(\bar x_i,\bar x_j) - \chi(x_{1i},x_{1j})
+\chi(x_{1i},x_{1j})] \\ &\sim
\log \left[ \int dx_{11}\cdots dx_{1p} \prod_{i<j}^{1,p} \chi(x_{1i},x_{1j})
+ \sum_{i<j}^{1,p} 
\int dx_{11}\cdots dx_{1p} \left(\prod_{i'<j'}^{1,p} \chi(x_{1i'},x_{1j'})\right)
Q(x_{1i}-x_{1j})
\right]
\ ,
\end{split}\eeq
where we omitted the second order in the development in series of $\bar\chi -
\chi_1$ and we defined
\beq
Q(x-y) = \int dx_1\cdots dx_m dy_1\cdots dy_m \r(\bar x) \r(\bar y) \left[
  \prod_{a=2}^m \chi(x_a,y_a) - 1 \right] \ . 
\eeq
In \cite{PZ10} it is shown that the second order gives a contribution $O(A)$
and that at lowest order (see Appendix C3 of \cite{PZ10}) $Q(r) =2 \sqrt{A}
Q_0(m) \d(r-D)$, where $Q_0(m)$ is a function of $m$ defined in
\cite{PZ10} as:
\beq
Q_0(m) = \int_{-\infty}^{\infty}\left[\Theta(t)^m-\Theta(t)\right]\ \ \ ; \ \
\Theta(t)=\frac{1}{2}[1+\text{erf}(t)]=
\frac{1}{\sqrt{\pi}}\int_{-\infty}^t dx e^{-x^2}
\eeq
We get then
\beq\begin{split}
S_{box} &\sim \log Z^0_p + \frac{p(p-1)}2 \int dx dy Q(x-y) g_0^p(x-y) \\ &=
\log Z^0_p + \frac{p(p-1)}2 \frac{2d \sqrt{A}}{D} Q_0(m) g_p^0(D) V_d(D)
\end{split}\eeq
and collecting all the terms we get
\beq\begin{split}
S(m) = &\frac{d}2 (m-1) \log (2 \pi A)
+\frac{d}2 \log m + \frac{d}2(1-m) (z-1) \log \left[ 1-\frac{1}z
\right] \\
&+ \frac{z}p \log Z^0_p + \frac{z (p-1)}2 \frac{2d \sqrt{A}}{D} Q_0(m) g_p^0(D)
V_d(D) \ .
\end{split}\eeq
Optimization with respect to $A$ gives
\beq\label{Astar_Gauss}
\sqrt{A^*} = D \frac{1-m}{Q_0(m)} \frac{1}{z (p-1) V_d(D) g^0_p(D)} \ ,
\eeq
and
\beq\label{SmGauss}
\begin{split}
S(m) = &\frac{d}2 (m-1) \log (2 \pi A^*)
+\frac{d}2 \log m + d(1-m)+ \frac{d}2(1-m) (z-1) \log \left[ 1-\frac{1}z
\right]
+ \frac{z}p \log Z^0_p \ .
\end{split}\eeq
In particular, using the results $Q_0(m\to 0) \sim \sqrt{\pi/4m}$ and
$Q_0(m \sim 1)=Q_0 \times (1-m)$ with $Q_0 = 0.638$~\cite{PZ10}, one can show
that this expression trivially reduces to the RS entropy (\ref{SRS}) for $m=1$,
and that
\beq\nonumber
\begin{split}
&\Si_j = \lim_{m\to 0} S(m) = 
-d \log \left[ \frac{2\sqrt{2} D}{z (p-1) V_d(D) g^0_p(D)} \right] + d +
\frac{d}2 (z-1) \log \left[ 1-\frac{1}z \right]+ \frac{z}p \log Z^0_p \ , \\
&\Si_{eq} = - \lim_{m\to 1} m^2 \partial_m [S(m)/m] =
- \frac{d}2 \log\frac{2\pi}{e} - d \log \left[ \frac{D}{z (p-1) V_d(D) g^0_p(D)
    Q_0}\right]
+  \frac{d}2 (z-1) \log \left[ 1-\frac{1}z \right] 
+ \frac{z}p \log Z^0_p
\end{split}\eeq

\begin{figure}[t]
\includegraphics[width=.9\textwidth]{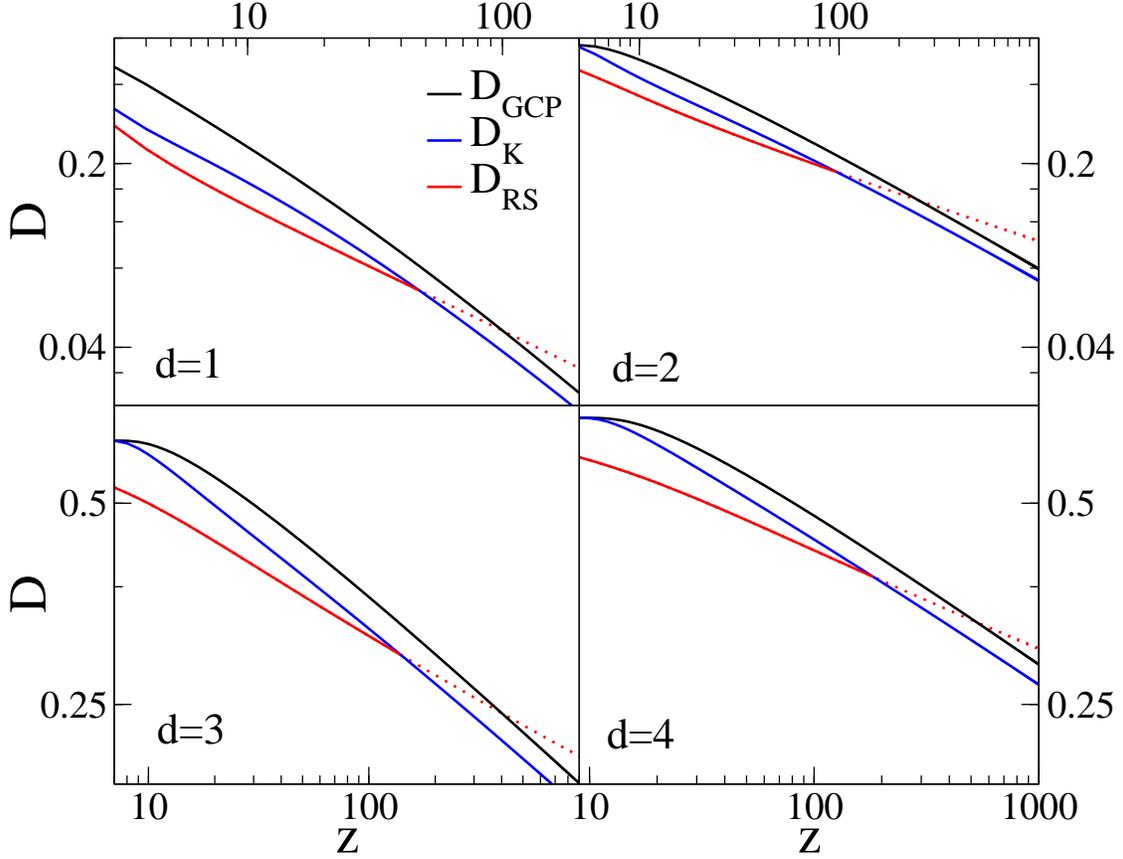}
\caption{Special values of the sphere radius as functions of
  $z$ at $p=2$ for different values of $d$ in the Gaussian approximation: $D_{RS}$ beyond which
  the RS solution becomes unstable, $D_{\rm GCP}$ where the pressure
  diverges,  and $D_{\rm K}$
  where the Kauzmann transition takes place. When $D_{\rm K}<D_{RS}$ the transition is necessarily first order.}
\label{gauss2}
\end{figure}

\begin{figure}[t]
\includegraphics[width=.9\textwidth]{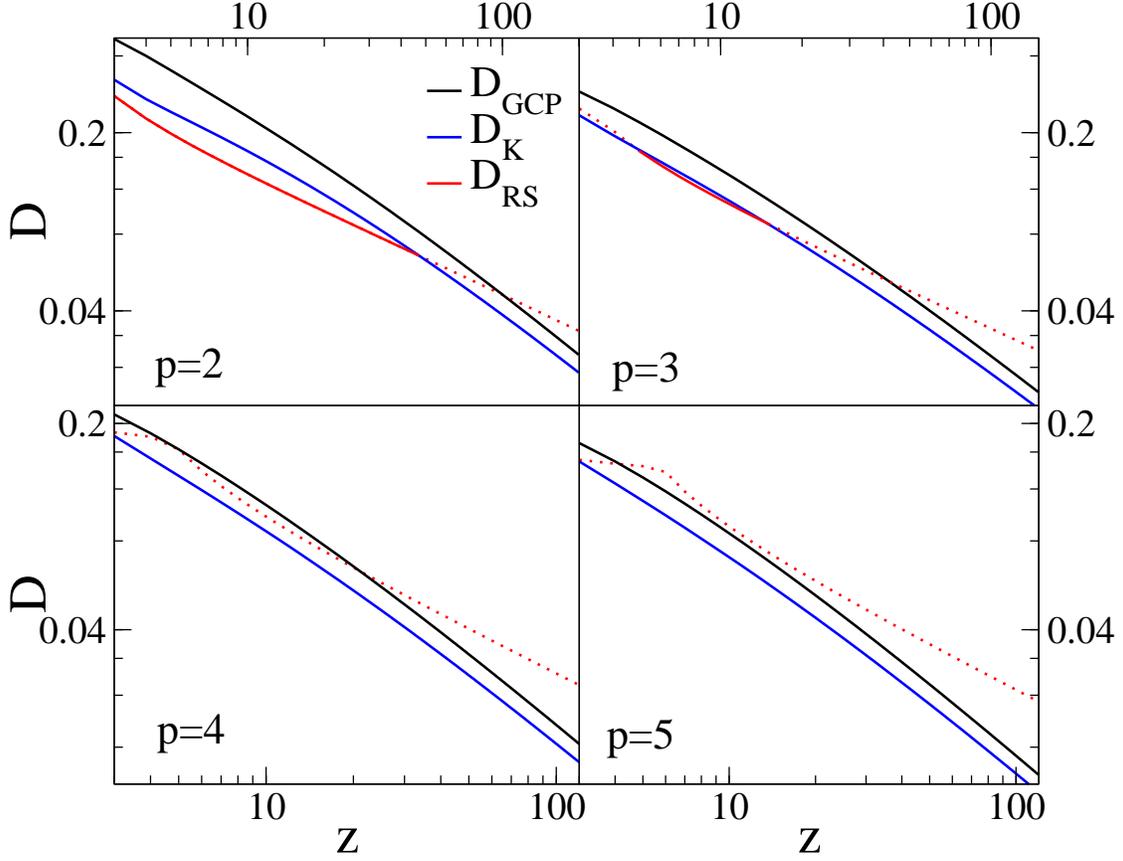}
\caption{$D_{RS}$, $D_{\rm GCP}$ and $D_{\rm K}$ as functions of
  $z$  for different values of $p$ at $d=1$ in the Gaussian approximation.}
\label{gauss}
\end{figure}

\subsection{Results for $p=2$, any dimension}

For $p=2$ we have trivially $Z^0_2 = 1 - V_d(D)$ and $g^0_2(x,y)=\chi(x,y)/Z^0_2$,
therefore $g^0_2(D)=1/Z_2^0$. We get
\beq\begin{split}
S(m) &= \frac{d}2 (m-1) \log \left[ \frac{2 \pi D^2 (1-V_d(D))^2}{z^2 V_d(D)^2} \frac{(1-m)^2}{Q_0(m)^2}\right]
+\frac{d}2 \log m \\ &+ d(1-m)+ \frac{d}2(1-m) (z-1) \log \left[ 1-\frac{1}z
\right] + \frac{z}2 \log [1-V_d(D)] \ ,
\end{split}\eeq
and
\beq\nonumber
\begin{split}
&\Si_j = \lim_{m\to 0} S(m) = 
-\frac{d}2 \log \left[ \frac{8 D^2 (1-V_d(D))^2}{z^2 V_d(D)^2} \right] + d +
\frac{d}2 (z-1) \log \left[ 1-\frac{1}z \right]+ \frac{z}2 \log [1-V_d(D)] \ , \\
&\Si_{eq} = - \lim_{m\to 1} m^2 \partial_m [S(m)/m] =
- \frac{d}2 \log \left[ \frac{2 \pi D^2 (1-V_d(D))^2}{z^2 V_d(D)^2
    Q_0^2}\right]
+\frac{d}2  +  \frac{d}2 (z-1) \log \left[ 1-\frac{1}z \right] 
+ \frac{z}2 \log [1-V_d(D)]
\end{split}\eeq
and $D_{\rm K}$ is defined by $\Si_{eq}=0$ while $D_{\rm GCP}$ is defined by $\Si_j=0$.
The results are reported in Fig.~\ref{gauss2}.

\subsection{Results for $d=1$, any $p$}

Also in $d=1$ the integrations can be performed for all $p$. We get
\beq\begin{split}
& Z^0_p = [1-p D]^{p-1}  \ , \\
& g^0_p(D) = \frac{1}{1-p D} \ .
\end{split}\eeq
Then
\beq\begin{split}
S(m) = &\frac{1}2 (m-1) \log \left[\frac{\pi (1-p D)^2}{4 z^2 (p-1)^2 } \frac{(1-m)^2}{Q_0(m)^2} \right] 
+\frac{1}2 \log m \\ & + (1-m)+ \frac{1}2(1-m) (z-1) \log \left[ 1-\frac{1}z
\right]
+ \frac{z(p-1)}p \log (1-p D) \ ,
\end{split}\eeq
and
\beq\nonumber
\begin{split}
&\Si_j = -
\frac{1}2 \log \left[\frac{2 (1-p D)^2}{z^2 (p-1)^2 } \right] + 1 + \frac{1}2 (z-1) \log \left[ 1-\frac{1}z \right]
+ \frac{z(p-1)}p \log (1-p D) \ , \\
&\Si_{eq} = 
- \frac{1}2 \log \left[ 
\frac{\pi (1-p D)^2}{2 z^2 (p-1)^2
    Q_0^2}\right]
+\frac{1}2  +  \frac{1}2 (z-1) \log \left[ 1-\frac{1}z \right] 
+ \frac{z(p-1)}p \log (1-p D) \ .
\end{split}\eeq
The results are reported in Fig.~\ref{gauss}.


\section{The delta approximation}

In this section we introduce 
another variational approximation scheme, that we shall call the ``delta approximation''.
The motivation is that within the Gaussian Ansatz, $A \to 0$ at jamming: therefore, 
both $\psi(x)$ and $\f(x)$ become delta functions in this limit.
We would therefore like to compute the free energy directly for delta function
fields; we expect this to give a simpler expression of the free energy, that should be good 
close to jamming.
The problem is that the Gaussian expressions are divergent for $A\to 0$ unless $m$
also goes to zero proportionally to $A$. 
This is due to the fact that {\it both} fields $\psi(x)$ and $\f(x)$
become delta functions for $A \to 0$. We therefore construct here a different approximation
by eliminating the field $\f(x)$ and making a delta function Ansatz only for the field $\psi(x)$:
in this way the field $\f(x)$ is computed exactly and in particular it is not a delta function.

One can show in general that by using equations (\ref{rec1RSB}), 
one can eliminate the field $\f(x)$ and
the replicated entropy can be equivalently written as
\beq
S(m) = S_{site'} - \frac{z(p-1)}{p} S_{box}
\eeq
where $S_{box}$ is defined as in Eq.~(\ref{S1RSB}) and
\beq\begin{split}
S_{site'} &= 
 \log \int d\PP[\psi^1_1]\cdots d\PP[\psi^z_{p-1}] 
\left[ \int dx \prod_{k=1}^z \int dx^k_1 \cdots dx^k_{p-1} \psi_1^k(x^k_1) \cdots \psi_{p-1}^k(x_{p-1}^k) 
\chi(x,x^k_1,\cdots,x^k_{p-1}) \right]^m \\
&\equiv \log \la Z_{site'}^m \ra \ .
\end{split}\eeq

The ``delta approximation''
is then based on the following Ansatz for $\PP(\psi)$:
\beq\label{PPdelta}
\PP[\psi] = \int d X \, \delta \left[ \psi(x) - \delta (x - X) \right] \ ,
\eeq
namely on each site $i$ the probability of the variable $x_i$ is a delta
function centered in a i.i.d. random point.
Under approximation (\ref{PPdelta}),
the replicated entropy becomes
\beq\label{Smdelta}
\begin{split}
S(m) &= \log \int dX_1^1 \cdots dX_{p-1}^z \left( \int dx 
\prod_{k=1}^z \chi(x,X^k_1,\cdots,X^k_{p-1}) \right)^m - 
\frac{z(p-1)}{p} \log \int dX_1 \cdots dX_p \chi(X_1,\cdots,X_p) \\
&= 
\log \int \left( 
\prod_{k=1}^z  dX_1^k \cdots dX_{p-1}^k \chi(X^k_1,\cdots,X^k_{p-1}) \right)
v_{z(p-1)}(X_1^1 \cdots X_{p-1}^z)^m
 - 
\frac{z(p-1)}{p} \log Z^0_p
\ ,
\end{split}\eeq
recalling the definition of $v_n$ in Eq.~(\ref{voidspace}).
Introducing the normalized measure of $n$ spheres
in a unit box,
\beq
d\mu(x_1 \cdots x_{n}) = \frac{dx_1 \cdots dx_{n} \chi(x_1 \cdots x_n)}{Z^0_{n}} \ ,
\eeq
we can rewrite $S(m)$ given in Eq.~(\ref{Smdelta})
in the equivalent form
\beq\label{Smdelta_sampling}
\begin{split}
S(m) = \log \int \left( \prod_{k=1}^z d\mu(X_1^k \cdots X_{p-1}^k) \right)
\left[ v_{z(p-1)}(X^1_1,\cdots,X^z_{p-1}) \right]^m 
+ z \log Z^0_{p-1}
 - \frac{z(p-1)}{p} \log Z^0_p \ .
\end{split}\eeq
In the following we study this expression for several specific values of $p$ and $d$.
In this section we will derive the expressions for the complexity, and in
section~\ref{sec:results} we will present the results together with a comparison with numerical resolution
of the cavity equations.
Note that for $m=1$ one can easily show that $S(m)$ given above is equal
to the RS entropy (\ref{SRS}), which is an important requirement for the consistency
of this approximation.

\subsection{One dimension}

\subsubsection{Results for $p=2$}

We first consider the simplest case, namely one spatial dimension and only 
two-particles-in-a-box interactions ($p=2$). Since $Z^0_1 = 1$ and $Z^0_2 = (1-2D)$, we get
\beq
S(m) =\log \int \prod_{i=1}^z dX_i \left[ v_z(X_1 \cdots X_z)
\right]^m - \frac{z}2 \log(1-2 D) \ .
\eeq
We have therefore to compute the probability distribution $P_z(v)$ of 
the void space left in $[0,1]$
for the insertion of a new particle, after having put $z$ particles in 
random positions $\{ X_i \}$. Then we have
\beq\label{SmPvapp}
S(m) =\log \int_0^{1-2D} dv \, P_z(v) \, v^m  - \frac{z}2 \log(1-2 D) \ .
\eeq
Note that $v$ ranges from $0$ (no void space) to $1-2D$ (in the limiting
case where all points $X_i$ coincide), and we expect that
$P_z(v) = p_0 \d(v) + P_z^{reg}(v)$ since a finite fraction of configurations have
zero void space at large enough $D$. Since the delta function does not
contribute to $S(m)$, we will omit it from now on.

In order to estimate $P_z(v)$ we can make the assumption that
whenever $v>0$, there is only one hole large enough to contribute to $v$ 
(\ie a hole whose length is bigger than $2D$). 
The function $P_z(v)$ can then be easily evaluated in the following way.
The hole that contributes to $v$ must have length $2 D + v$, and must
be delimited by two particles that we can choose in $z (z-1)$ different ways, 
since particles are distinguishable. We can put the first particle in
$x_1=0$ and the second in $x_2 = 2 D + v$ 
(integration over $x_1$ can be omitted since it gives a factor of 1, the length
of the box).
The remaining $z-2$ particles must be in the space between $x_2$ and $1$,
therefore giving a contribution $(1 - 2 D - v)^{z-2}$.
Therefore, within the one-hole approximation, we get 
$P_z(v) = z (z-1) (1-2 D -v)^{z-2}$. 
We notice that the total probability of $v>0$ 
must be smaller then one since some configurations might have $v=0$.
This gives the condition
\beq\label{one-hole-app-p2}
\int_0^{1-2D} dv P_z(v) = z (1-2 D)^{z-1} \leq 1
\hskip1cm \Rightarrow \hskip1cm  D \geq (1 - z^{-1/(z-1)})/2 \ ,
\eeq
which gives an estimate
of the limits of validity of the one-hole approximation.

Plugging the result for $P_z(v)$ in Eq.~(\ref{SmPvapp}), we get
an approximate formula for the replicated free energy which 
depends on $z$ and $D$,
\beq
S(m) = \log\left(\frac{\Gamma(z+1)\Gamma(m+1)}{\Gamma(z+m)}\right) +\left( m - 1 + \frac{z}{2} \right)\log (1 - 2D) \ .
\eeq 
Recall that $\Si_{eq} = -[m^2 \partial_m (S(m)/m) ]\vert_{m=1}$ and that
$D_{\rm K}$ is the point where the latter quantity vanishes.
We get
\beq\label{eq_delta_d1_p2}
\Si_{eq} = 
 \sum_{q=2}^{z}\frac{1}{q}
+\frac{z-2}{2}\log\left({1-2 D}\right)\ , 
\hskip1cm
D_{\rm K}=\frac{1}{2}\left[ 1-e^{-\frac{2}{z-2}\sum_{q=2}^{z} \frac1q }\right] \ .
\eeq
On the other hand, $\Si_j = S(m=0)$ and it vanishes at the
close packing diameter $D_{\rm GCP}$. We get
\beq\label{j_delta_d1_p2}
\Si_j =\log(z) + \frac{z-2}{2}\log(1-2 D) \ ,
\hskip1cm
D_{\rm GCP}=\frac{1}{2}\left[1-z^{-2/(z-2)}\right] \ .
\eeq
The complexity curve can be obtained explicitely, using $\Sigma = - m^2 \partial_m (S(m)/m)$ and $s=\partial_m S(m)$, 
which gives the parametric representation:
\beq\begin{split}
& s=\log(1-2 D)-\sum_{q=1}^{z-1}\frac{1}{m+q}\ , \\
& \Sigma=\frac{z-2}{2}\log(1-2 D)+
\log\left( \frac{\Gamma(z+1)\Gamma(m+1)}{\Gamma(z+m)} \right)
+m \sum_{q=1}^{z-1}\frac{1}{m+q}\ .
\end{split}\eeq

One can check easily that
both critical diameters $D_{\rm K}$ and $D_{\rm GCP}$ are well 
within the region of validity of the one-hole
approximation given by Eq.~(\ref{one-hole-app-p2}), and they
scale as 
$D_{\rm K},D_{\rm GCP} \sim \log z / z$ in the large connectivity
limit.
The values of $D_{\rm K}$ and $D_{\rm GCP}$ can be compared to the stability of the RS 
solution (which scales as $D_s \sim 1/\sqrt{z}$). 

\subsubsection{Results for $p=3$}

We now consider the three-particles-in-a-box case $p=3$, still for $d=1$. 
Since $Z^0_2 = 1-2D$ and $Z^0_3 = (1-3 D)^2$, we get from
Eq.~(\ref{Smdelta_sampling}):
\beq\label{SmPvapp_p3}
S(m) =\log \int_0^{1-3D} dv \, P_{2,z}(v) \, v^m  + z \log(1-2 D) - \frac{4z}{3} \log(1-3D) \ .
\eeq
where now $P_{2,z}(v)$ is the probability distribution of the void space in $[0,1]$
for the insertion of a new particle, after having thrown at random $z$ 
pairs of particles, each pair being at distance bigger than $D$.
The latter ranges from $0$ (no void space) to $1-3D$ (in the case where
each pair is exactly at distance $D$ and superposed to all the others).
 
Within the same one-hole approximation, 
we can approximate $P_{2,z}(v)$ as follows. 
The hole must have length $L=2 D +v$.
We have to distinguish 
between two different situations: {\it i)} The hole is made by the same
couple of particle; {\it ii)} The hole is made by two different couples.
In the case {\it i)} we have $z$ ways of choosing the couple. We fix then one of 
the two particles of the couple in $0$ and the other one in $L$ (which 
gives an extra factor $2$). Finally the other $z-1$ couples of particles
must be in the interval $[L,1]$ with the conditions that they are pairwise 
compatible, which gives a factor 
$f(L,D) = \int_L^1 dx \int_L^1 dy \, \chi(x,y) = (1 - L - D)^2$ for each pair.
With this definition the contribution due to the same couple finally reads:
$2 z f(L,D)^{z-1}$. 
In the case {\it ii)}, instead, we can fix one particle of one couples in $0$
(we have $2 z$ ways to choose it) and one particle of another couple in $L$
(we have $2 (z-1)$ ways of choosing it). The free particle of the first couple
must be in $[L,1-D]$, due to 
the condition that it is compatible with its partner
which has been fixed in $0$. This gives a contribution $(1 - L - D)$. 
An analogous contribution comes from the the free particle of the second 
couple, which must be in the interval $[L+D,1]$. 
The other $z-2$ couples must be in the interval $[L,1]$
and must satisfy the compatibility condition, and therefore give a contribution
$f(L,D)^{z-2}$. The sum of the two contributions is
$(4 z^2 - 2 z) (1 - L - D)^{2(z-1)}$, and it has to be normalized by the 
total integral $(1-2 D)^z$; going back to $v = L -2D$ we get
\beq
P_{2,z}(v) = \frac{2 z (2 z - 1) (1 - v -3 D)^{2(z-1)}}{(1-2D)^z} \ .
\eeq
As in the previous case we get the condition
\beq\label{one-hole-app-p3}
\int_0^{1-3D} dv \, P_{2,z}(v) = \frac{2z (1-3 D)^{2z-1}}{(1-2D)^z} \leq 1 \ ,
\eeq
which gives a lower limit of validity in $D$ of the one-hole approximation.

Plugging this results in Eq.~(\ref{SmPvapp_p3})
we get for the replicated entropy
\beq\label{eq_delta_d1_p3}
S(m) = \log \left[ 
\frac{ \Gamma(m+1) \Gamma(1 + 2 z) }
{ \Gamma(m + 2 z) }
 \right] + 
\left( m-1 - \frac{2z}{3} \right) \log(1-3 D)
\eeq
from which we get
\beq\label{j_delta_d1_p3}
\Si_{eq} = \sum_{q=2}^{2 z} \frac{1}{q}+ \frac{2 z-3}3  \log(1 - 3 D)
\ , \hskip1cm D_{\rm K} = \frac13 \left[ 1 -e^{-\frac{3}{2z-3}\sum_{q=2}^{2 z} \frac{1}{q}
} \right] \ ,
\eeq
and
\beq
\Si_j = \log(2 z) + \frac{2z-3}3 \log(1 - 3 D) 
\hskip1cm
D_{\rm GCP}=\frac{1}{3}\left[ 1-(2z)^{-3/(2z-3)}\right]\ .
\eeq
We checked that both $D_{\rm GCP}$ and $D_{\rm K}$ are well within the region
of validity of the one-hole approximation; actually, the value of the left hand side
of Eq.~(\ref{one-hole-app-p3}) never exceeds 0.1.
Again, $D_{\rm GCP}$ and $D_{\rm K}$
are found to scale as $2 \log z /z$ for large $z$.

\subsubsection{Conjecture for arbitrary $p$ ($2,3,\cdots,\io$)}

A comparison of Eqs.~(\ref{eq_delta_d1_p2}) and (\ref{eq_delta_d1_p3})
and of Eqs.~(\ref{j_delta_d1_p2}) and (\ref{j_delta_d1_p3})
allows to guess the form for general $p$:
\beq\begin{split}
& \Si_{eq} = \sum_{q=2}^{(p-1) z} \frac{1}{q}+ \frac{(p-1) z-p}p  \log(1 - p D)
\ , \hskip1cm D_{\rm K} = \frac1p \left[ 1 -e^{-\frac{p}{(p-1)z-p}\sum_{q=2}^{(p-1) z} \frac{1}{q}
} \right] \ ,\\
&
\Si_j = \log((p-1) z) + \frac{(p-1)z-p}p \log(1 - p D)  \ ,
\hskip1cm
D_{\rm GCP}=\frac{1}{p}\left[1-((p-1)z)^{-p/((p-1)z-p)}\right]\ .
\end{split}\eeq
however we did not attempt to provide a proof of this conjecture.

\subsection{Two dimensions}
\label{sec:deltad2}

In the $d=2$ case we cannot compute $S(m)$ analytically and
we must resort to a numerical evaluation.
The numerical algorithm consists in writing a routine that is able to compute
the void space $v_n$, defined in Eq.~(\ref{voidspace}), 
left by $n$ disks centered in a set of positions $\{X\}$. 
We used an adaptation of the algorithm described in \cite{RT95} that
works as follows:
\begin{itemize}
\item We start by a grid of squares of side $\D \ll D$ (typically $\D = 1/100$).
These squares are considered as particular cases of convex polygons.
\item We add disks $X_1 \cdots X_n$ sequentially.
\item Each time a disk is added, we check if a given polygon is entirely 
contained in the disk. In this case it is removed from the grid.
\item Next we consider the polygons that intersect the boundary of the new disk.
We approximate the boundary of the void space left in the old polygon by a 
new polygon, by approximating the boundary of the disk by a straight line (which
is reasonable if $\D \ll D$, with error $O(\D/D)^2$). The new polygon replaces the old one in the grid.
\item This construction is iterated until all disks have been placed. The area
of the polygons that survived is computed easily using Eq.~(1) of Ref.~\cite{RT95},
and it gives the void space $v_n$.
\end{itemize}

The void space has to be averaged over the 
distribution $\prod_{i=1}^z d\mu(X^i_1 \cdots dX^i_{p-1})$, 
hence we must sample a configuration of $p-1$ spheres in a box
(and do this $z$ times indepentently).
This can be easily done for $p=2$ (one sphere, flat distribution) and
$p=3$ (put one sphere in the centre of the box, draw a second sphere outside
it, then translate randomly both spheres). 

A correct sampling gives access to the void space distribution $P(v)$,
that has the form $P(v) = p_0 \d(v) + P^{reg}(v)$, as in one dimension. 
In the following we omit the delta term and only consider $P^{reg}(v)$,
which therefore is not normalized to one (its integral gives the probability
that $v>0$).
From this we can compute Eq.~(\ref{Smdelta_sampling}) as we did in one dimension:
\beq
S(m) = \log \int dv \, P(v) \, v^m 
+ z \log Z^0_{p-1}
 - \frac{z(p-1)}{p} \log Z^0_p \ .
\eeq
Similarly we get, using the relation 
$\int dv \, P(v) \, v = \la v \ra = (Z^0_p/Z^0_{p-1})^z$ (which can be easily checked
and also serves as a check of the correct sampling of $P(v)$),
\beq\begin{split}
\Si_{eq} &= \frac{z}{p} \log Z^0_p - \left( \frac{Z^0_{p-1}}{Z^0_p} \right)^z \int dv \, P(v) \, v \log v \ , \\
\Si_j &= \log \int dv \, P(v) + z \log Z^0_{p-1}
 - \frac{z(p-1)}{p} \log Z^0_p \ .
\end{split}\eeq
Therefore both $\Si_{eq}$ and $\Si_j$ can be computed directly from $P(v)$; from them we can determine
the transition points $D_{\rm K}$ and $D_{\rm GCP}$.


\section{Numerical solution of the equations}

In the previous sections we described two analytical approximate
methods yielding the phase diagram of the model.
Beyond these analytical approaches, one can also develop some 
algorithms to solve the functional self-consistent 1RSB equations
numerically.
In this section we explain how it is possible to implement a numerical
procedure to solve Eqs.~(\ref{rec1RSB}) in the 1RSB phase for
each value of the connectivities, $z$ and $p$, of the diameter $D$, of
the 1RSB parameter $m$ and, in principle, of the spatial dimension
$d$ (in practice, numerical solutions can only be achieved in one and two
dimensions).
In order to do that we need representations of the cavity fields
$\f(x)$ and $\psi(x)$, and of the distributions
$\cal{P} [\f]$ and $\cal{P} [\psi]$, which can be treated by a computer.

As far as the cavity fields are concerned, 
the simplest possibility is to discretize the volume $[0,1]^d$ where the
functions $\f(x)$ and $\psi(x)$ are defined using a regular hyper-cubic
grid with $q$ bins per side of size $1/q$. For instance, in one dimension 
we discretize the interval $[0,1]$ in $q$ slices of length $1/q$, and
in two dimension we discretize the square box on a square lattice of
$q\times q$ points. 

The coordinate in the
box can assume a discrete set of values, $\vec{i}/q$, with $\vec{i}$ being
a $d$-dimensional vector whose components are integers between $0$ and 
$q-1$, identifying the coordinate of the 
position of the center of the sphere in the box.
If the position of the center of the sphere occupies a given site of the grid
$\vec{i}$, then all other sites of the lattice that are at Euclidean distance 
from $\vec{i}$ smaller than the diameter of the sphere $D$
cannot be occupied by the center of another sphere (we call this number $n_D$). 
The volume of the sphere in the discretized version of the model can 
be estimated as $V_s = n_D/(2q)^d$, and the packing fraction as 
$\f = p V_s = p n_D/(2q)^d$.
Since in the continuum limit $V_s = V_d(1) (D/2)^d$,
we can then define an effective diameter as
$D_{\rm eff} = \frac1q \left[\frac{n_D}{V_d(1)} \right]^{1/d}$.
Note that in general $D_{\rm eff} \neq D$, and we take $D_{\rm eff}$ as representative
of the sphere diameter in the continuum limit. In particular, by symmetry, 
in $d=1$ the number of excluded
sites always has the form $n_D = 1 + 2 a$ for integer $a$, 
and one has 
\beq\label{effDd1}
D_{\rm eff} = \frac{1+2 a}{2q} \ .
\eeq
In $d=2$ the parameter $n_D$ depends in an irregular manner on the choice of $D$
(since the square lattice we use breaks the spherical symmetry) and one has in general
\beq\label{effDd2}
D_{\rm eff} = \frac1q \sqrt{ \frac{n_D}{\pi} }  \ .
\eeq

In the discretized version, the fields $\f(x)$ and $\psi(x)$ are  
vectors of $q^d$ components
(such that the sum of all components is equal to one), and
the cavity equations, Eqs.~(\ref{recurrence}), become
a set of coupled algebraic equations for the $q^d$ components
of the cavity fields, which can be easily solved numerically
(of course, the numerical complexity of this step grows linearily with 
the number of components of the cavity fields, $q^d$).

Note that the discretized version of the model is a generalization of
a very important optimization problem known as the ``random graph coloring'' 
problem, where the number of colors corresponds
to the number of components of the cavity fields $q^d$. In particular, 
for $n_D = 0$ and $p=2$ we recover 
the standard $q$-coloring problem, which has been deeply
studied in the past few years, and whose properties and phase diagram are 
known in great details~\cite{col2}.

The continuum limit of the model is, of course, recovered 
for $q \to \infty$. As a consequence, in order to make sure that
the numerical results are reliable and that they are not
affected by the discretization, 
we solve numerically the 1RSB equations using several values 
of $q$, and analyze the scaling properties of the numerical solutions 
with the number of bins.
Moreover, one should note that for $d>1$, partitioning the box using an 
hyper-cubic grid breaks the spherical symmetry down to some discrete
symmetry. This makes the scaling
towards the continuum limit in two dimensions more problematic than in one 
dimension (also
because, due to the fact that the complexity of the numerical algorithm 
grows as $q^d$, we are limited to smaller values of $q$ for $d=2$).

Other numerical representations of the cavity fields were also
possible. For instance, as $\f(x)$ and $\psi(x)$ are periodic functions
in the interval $[0,1]^d$, we could have performed a Fourier transformation
of the recurrence equations keeping all the components up to
a certain momentum, yielding a finite set of coupled algebraic 
equations for the Fourier coefficients of the cavity fields (similarily
to what we did in Sec.~\ref{sec:RSstability} to study the RS stability).
However, it turns out that this strategy is not efficient in the most
interesting region of the phase diagram, namely at high packing fraction
where a 1RSB glass transition is found. Indeed here the cavity fields
becomes extremely peaked (this is also the reason why the Gaussian and the
delta approximation work very well), and the momentum cut-off needed
to get accurate results becomes too big to be handled.

Another possibility we could have employed, is to represent the 
fields as a population of delta functions, \eg $\f (x) = \sum_\alpha
c_\alpha \delta (x - x_\alpha)$. This strategy, which has the advantage 
that one does not need to discretize the space, has, on the other hand, the 
disadvantage that at each step of the iterative procedure, in order to generate
a new field, one has to sample uniformly one point in
the free space available for the insertion of a new particle, given the 
position
of $z(p-1)$ neighboring particles in the box. 
This is trivial in $d=1$, however in that case the discretized procedure
work already well enough.
In $d=2$, this could be done using the
algorithm described in Sec.~\ref{sec:deltad2}. However 
this algorithm is too slow to be used efficiently to this scope. Therefore
in the following we will not explore further this representation.

\subsection{The population dynamics algorithm}

Now, once that we dispose of the discretized representation of the cavity
fields, we need to be able to implement a computational strategy to
solve the 1RSB
functional self-consistent equations, Eqs.~(\ref{rec1RSB}), for any value
of the connectivities, $z$ and $p$, of the diameter of the spheres, $D$, and 
of the 1RSB parameter $m$.
This step is quite standard in the context of the cavity 
method, and goes 
under the name of ``population dynamics algorithm''~\cite{cavity}.
The idea is to represent the probability distributions $\cal{P} [\f]$ 
and $\cal{P} [\psi]$ as populations of $\cal{M}$ representative 
cavity fields with some weights:
\begin{equation}
{\cal P} [\f] = \sum_{\alpha = 1}^{\cal M} z_\f^\alpha \, \delta [ \f (x)
- \f_\alpha (x) ] , \qquad \textrm {and} \qquad
{\cal P} [\psi] = \sum_{\alpha = 1}^{\cal M} z_\psi^\alpha \, \delta [ \psi (x)
- \psi_\alpha (x) ]
\end{equation}
As previously discussed, we need to consider only translationally invariant
solution of Eqs.~(\ref{rec1RSB}) in order to describe the glassy phase.
A solution $\PP[\psi(x)]$ is translationally invariant if the property
$\PP[\psi(x+s)] = \PP[\psi(x)]$ holds for any $s \in [0,1]^d$, where
$\psi(x+s)$ is an arbitrary translation (taking into account periodic
boundary conditions) of $\psi(x)$.
Since we represent the probability distribution $\PP[\psi]$
by a set of representative samples $\psi_\a(x)$, it is very
easy to implement translational invariance. In principle, we would like
to impose that if $\psi_\a(x)$ is one of the samples, then any translation of it
is also contained in the set of samples with the same weight. But this is just equivalent to do the
following: at each time we use a given sample $\psi(x)$ as a representative
of $\PP[\psi]$, we apply to it a ``random shift'', namely we extract a vector
$s$ uniformly in $[0,1]^d$ and we translate $\psi(x)$ by $s$. In this way
we impose translational invariance by hand.

The population dynamics algorithm works in the following way:
\begin{itemize}
\item[1)] Pick at random $p-1$ fields $\psi_i$ from the population 
$\cal{P} [\psi]$, according to their weights $z^\alpha_\psi$.
Apply a random
shift with flat probability in $[0,1]^d$ to each of the cavity fields.
\item[2)] Using Eq.~(\ref{recurrence}), compute the new cavity field
$\f$, along with its weight $z_\f$, which is given by the normalization
in Eq.~(\ref{reweighting}) to the power $m$, according to Eq.~(\ref{rec1RSB}). 
Note that at high density, in the
1RSB phase, the cavity fields becomes extremely peaked. This implies that
there exist some configurations of the $p-1$ fields $\psi_i$ for which
the new field $\f$ is zero everywhere in $[0,1]^d$.
In this case the corresponding weight is zero and we have to reject it
and restart the procedure. 
These events, which can cause a 
major slowing down of the algorithm, are called ``rejection events''.

\item[3)] Repeat 1) and 2) ${\cal M}$ times, until a whole new population
${\cal P}_{new} [\f]$ is generated, and replace the old population
with the new one (this kind of update is called in the context of population
dynamics algorithm ``parallel update'').
\item[4)] Apply steps 1), 2), and 3) using the population 
${\cal P} [\f]$ to generate a new ${\cal P}_{new} [\psi]$.
\item[5)] Repeat steps 1), 2), 3), and 4) until convergence, namely 
until the populations ${\cal P} [\psi]$ and ${\cal P} [\f]$ are
stationary.
\end{itemize}
Once this process has converged, we can compute the average values of 
the link, the site and the box contribution to the 1RSB entropy,
Eq.~(\ref{S1RSB}), from which one can obtain the complexity $\Sigma(m)$.
This allows to determine the equilibrium value of $m^{\star}$ 
inside the 1RSB glassy phase 
as the point where $S(m)$ has a minimum~\cite{Mo95}.
In practice, instead of computing the replicated entropy using 
Eq.~(\ref{S1RSB}), we can use another and equivalent formula (derived below) 
which is more advantageous from a numerical point of view.
Indeed, using Eqs.~(\ref{recurrence}) we can easily obtain
the following relations (we omit the arguments of the functions $Z$):
\begin{equation}
Z_{link} = \frac{ Z_{box}}{Z_\f} = \frac{ Z_{site}}{Z_\psi} \ .
\end{equation}
Using these and Eqs.~(\ref{rec1RSB}),
one can rewrite the total and internal entropy as
\begin{eqnarray}\label{S1RSBnv}
\nonumber
S(m) &=& 
\left(1-z+\frac{z}p\right) S_{link} + \frac{z}p S_{\f} + S_{\psi} \\
s(m) &=& 
\left(1-z+\frac{z}p\right) s_{link} + \frac{z}p s_{\f} + s_{\psi}
\end{eqnarray}
The computation of 
$S_{\f} = \log \langle Z_\f^m \rangle$ and $S_{\psi} = \log \langle Z_\psi^m 
\rangle$ is numerically less involved than
$S_{site}$ and $S_{box}$ appearing in Eq.~(\ref{S1RSB}). Moreover,
these contributions can be evaluated on-line during steps 1)-5) of the 
population dynamics algorithm described above (we have just to compute the 
average value of $Z_\f^m$ and $Z_\psi^m$ over all the ${\cal M}$ attempts of 
generating a new cavity field), without requiring the implementation of any 
further step.

Of course, representing the distributions $\cal{P} [\psi]$ and $\cal{P} [\f]$
as populations of ${\cal M}$ elements is an approximation which becomes
exact only in the ${\cal M} \to \infty$ limit. On the other hand, the numerical
complexity of the population dynamics algorithm grows linearily with
${\cal M}$. 
In practice on has to find a good compromise between a value of ${\cal M}$
small enough such that the execution time of the code stays reasonable,
but big enough to avoid systematic corrections due to the finite size
of the populations.
In the present case, we find that ${\cal M} = 2^{16}$ is close to the
optimal value.

Although we have produced a working version of the algorithm described above 
at any finite value of the 1RSB parameter $m$, it turned out that the
execution time is too big to get accurate results in a reasonable time.
However, there are two special limits, namely $m \to 1$ and $m \to 0$,
which describe respectively the physics at the Kauzmann point and 
in the close packing regime, where some semplifications arise which allow to 
perform the numerical study of the model in a more efficient way.
These two limits are discussed below.

\subsection{Reconstruction: the limit $m=1$}

In this section we consider the numerical solution of the 1RSB equations
for $m=1$. Recall that $S(m=1)$ gives back the equilibrium RS entropy of the system
between the dynamical transition (where a non-RS solution of the 1RSB 
equations 
appears for the first time due to the emergence of glassy metastable states)
and the Kauzmann point. 
In this limit, using the approach introduced 
in~\cite{MM06} which goes under the name of reconstruction method, 
also applied in a similar context to the coloring optimization problem 
in~\cite{col2}, the self-consistenf 1RSB equations can
be simplified.
Similarily to~\cite{MM06,col2}, 
one can indeed introduce two new families of distributions over the
cavity fields for each value of the variable $x$, defined as
\begin{equation}
{\cal R}_x [\psi] \equiv \psi(x) {\cal P} [\psi] \qquad \textrm{and} \qquad
{\cal R}_x [\f] \equiv \f(x) {\cal P} [\f] \ .
\end{equation}
Using the previous definitions, the 1RSB cavity equations, 
Eqs.~(\ref{rec1RSB}) can be rewritten in terms of these new distributions.
Furthermore, imposing the translational invariance which implies
that ${\cal R}_x [\psi (y)] = {\cal R}_0 [\psi (y - x)]$
for all $x$ we obtain the
the self-consistent recursion relation
for the new distributions which read:
\begin{eqnarray}
{\cal R}_0 [ \psi ] &=& \int \prod_{i=1}^{z-1} d {\cal R}_0 [ \f_i ] \,
\d\left[ \psi(x) - \frac{1}{Z_\psi} \prod_{i} \f_{i}(x) \right] \\ 
\nonumber
{\cal R}_0 [ \f ] &=& \int d\mu(x_1 \cdots x_{p-1} | 0)
\prod_{i=1}^{p-1} 
d{\cal R}_{0} [ \psi_i ] \,
\d\left[ \f(y) - \frac{1}{Z_\f} \int 
\prod_j dy_j \psi_j(y_j-x_j) \chi(y,y_1,\cdots,y_{p-1}) \right]
\end{eqnarray}
where
\beq
 d\mu(x_1 \cdots x_{p-1} | 0) =
\frac{\chi (0, x_1, 
\cdots , x_{p-1}) dx_1 \cdots dx_{p-1}}{Z_p^0}
\eeq
From a numerical point of view, these latter equations are much easier
to solve than Eqs.~(\ref{rec1RSB}) for two reasons. First, no
reweighting factor is present, which prevent the population to concentrate
on few cavity fields with large weight. Second, 
rejection events cannot occur in this case. Indeed, for example, the 
procedure to generate a new field $\f$ amounts to:
\begin{itemize}
\item[1)] Pick at random $p-1$ fields $\psi_i$ from the population
${\cal R}_0 [\psi]$. Note that all the fields have the same weight in this
representation.
\item[2)] Pick $p-1$ variables $x_1, \cdots, x_{p-1}$ in the interval 
$[0,1]^d$ satisfing the hard-sphere constraint $\chi (0, x_1, 
\cdots , x_{p-1})$ with a flat measure.
\item[3)] Shift each of the $p-1$ chosen cavity fields $\psi_i$ by $x_i$.
\item[4)] Using Eq.~(\ref{recurrence}), compute the new cavity fields
$\f$ (again, note that there is no reweighting in this case), and 
insert the new field randomly into the population ${\cal R}_0 [\f]$
(this kind of update is called ``serial update'' and ensures a better
convergence than the parallel one).
\end{itemize}
Once the populations ${\cal R}_0 [\f]$ and ${\cal R}_0 [\psi]$  have
attained stationarity, we can compute the complexity of the system.
Since the replicated entropy $S(m=1)$ equals the RS one, 
the complexity at $m=1$ is given by $\Sigma_{eq} = S_{RS} - s(m=1)$.
The internal entropy can be evaluated using Eqs.~(\ref{sSig1RSB})
and (\ref{rec1RSB}), where
\begin{eqnarray}
\nonumber
\langle Z_{link} \log Z_{link} \rangle & = &
\int d {\cal R}_0 [\psi] d {\cal R}_0 [\f] \, \log \int dy \, \psi(y) \f (y) \\
\langle Z_{\psi} \log Z_{\psi} \rangle & = &
\int \prod_{i=1}^{z-1} d {\cal R}_0 [\f_i] \, \log \int dy \prod_i \f(y) \\
\nonumber
\langle Z_{\f} \log Z_{\f} \rangle & = &
\int d\mu(x_1 \cdots x_{p-1} | 0) Z_p^0
\prod_{i=1}^{p-1} d{\cal R}_{0} [ \psi_i ] \,
\log \int dy \prod_i d y_i \psi_i (y_i - x_i) \, \chi(y,y_1,\ldots, y_{p-1}). 
\end{eqnarray} 
From the complexity we can determine the Kauzmann point, 
which corresponds to the 
value $D_K$ where $\Sigma_{eq}$ vanishes.

In principle this method would also allow to determine the location of the
dynamical transition, which is the first point where a non-RS solution
of the 1RSB equations appear at $m=1$. 

The results at $m=1$ obtained with the reconstruction method will be discussed
in Sec.~\ref{sec:results}, and compared with the analytical approximations.

\subsection{Hard fields: the limit $m=0$}

Also this specific limit yields a simplification of the numerical algorithm.
The $m \to 0$ limit corresponds in this context to the
``close packing limit'', since an inspection of the expression of the internal
entropy $s(m)$ shows that it goes to $-\io$ as $\log(m)$, and the
pressure diverges as well~\cite{PZ10}.
Therefore the limit $m\to 0$ gives access to the jammed
glassy states at infinite pressure~\cite{PZ10}.

The limit for $m$ going to zero of $Z_{link}^m$, 
$Z_{box}^m$, and $Z_{site}^m$ are either zero (for ``incompatible'' 
configurations of the cavity fields) 
or one (for ``compatible'' configurations of the cavity fields) 
regardless of the value of the cavity fields.
As a consequence, in order to compute the complexity 
(which equals the replicated entropy $S(m \to 0)$, since the internal 
entropy term, $m s(m)$, disappears) we are only 
interested in the propagation of this information.

To this aim, we introduce the ``hard'' components of the cavity fields
$\psi_{hard}$ and $\f_{hard}$:
\begin{equation}
\psi_{hard} (x) = 
\left \{
\begin{array}{ll}
1 & \textrm{if $\psi(x)>0$}\\
0 & \textrm{otherwise}
\end{array}
\right.
\qquad \textrm{and} \qquad
\f_{hard} (x) = 
\left \{
\begin{array}{ll}
1 & \textrm{if $\f(x)>0$}\\
0 & \textrm{otherwise}
\end{array}
\right.
\end{equation}
These functions are defined as being equal to one for all values of
$x$ such that the cavity fields are non vanishing regardless of their
value (i.e., corresponding to a non-vanishing probability of finding a sphere
with center in $x$), and zero otherwise.
Since the reweighting factors in Eq.~(\ref{rec1RSB}) do not depend on
the actual value of the fields in the $m \to 0$ limit, 
the propagation of the hard components
decouples completely from the propagation of the cavity fields and can thus
be treated indepenently. As a consequence, the population dynamics algorithm
described above can be used on the populations encoding the 
probability distributoons of the hard fields.
Once a stationary state has been reached, we can compute the complexity 
at $m=0$, $\Sigma_j$, from Eq.~(\ref{S1RSB}), computing the logarithm of the
average value of the fraction of attempts yielding a non vanishing
value of $Z_{link}$, $Z_{box}$, and $Z_{site}$.
Using Eq.~(\ref{S1RSBnv}), instead of computing  
$\langle Z_{box}^m \rangle$ and $\langle Z_{site}^m \rangle$, 
one can more easily compute $\langle Z_{\psi}^m \rangle$ and 
$\langle Z_{\f}^m \rangle$, which are given respectively by the average value 
of the fraction of non-rejection attempts to generate the new $\psi_{hard}$ and 
$\f_{hard}$ fields over the total number of attempts.
Then we can determine the location of $D_{GCP}$ defined as $\Sigma_j
(D_{GCP}) = 0$.

The results at $m=0$ obtained with this method will be reported
in Sec.~\ref{sec:results}, and compared with the analytical approximations.

An important {\it caveat} is that in principle some fields could be proportional
to $\exp(-1/m)$ in the limit $m\to 0$. If this happens, then the procedure above
fails since these fields give a finite contribution to the normalizations which
is neither 0 nor 1. Although we could not perform a careful systematic investigation
of this effect, it seems that it might happen only for values of $z$ and $p$ 
where the transition at $m=1$ is continuous. This point surely deserves further 
investigation.

Note that 
in order to compute the correlation function in the close packing limit
(see Sec.~\ref{corr-func}) we also need to know the 
actual values of the
cavity fields. Since the propagation of the hard components decouples
completely from the the one of the fields itself, one can use the population
dynamics algorithm to find the solution of the 1RSB equations for
the distributions of hard fields and of the cavity fields independently
(knowing that the cavity fields can only be non zero where the 
hard components are equal to one),
and use Eq.~(\ref{eq:gr}) to compute the pair correlation function.


\begin{figure}[t]
\includegraphics[width=.9\textwidth]{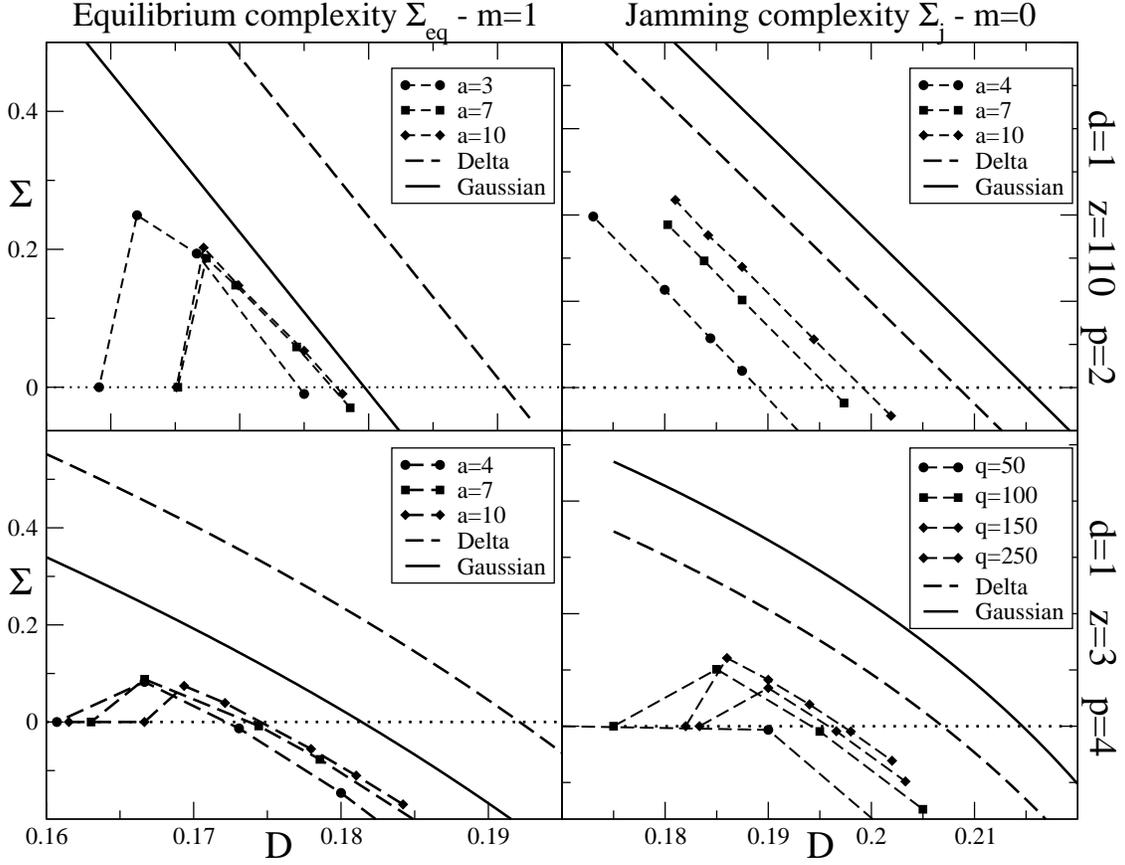}
\caption{
The complexity in some representative cases of discontinuous
transition at $d=1$, computed with the numerical solution of the
population dynamics algorithm with varying resolution of the
discretization process, is compared to the Gaussian and the Delta approximations. 
[Upper panels] $\Si_{eq}$ (left) and
$\Si_j$ (right) for
$p=2$ and $z=110$. In both cases 
we fixed the parameter $a = 4,7,10$ in Eq.~(\ref{effDd1}) and changed $q$ to vary
the effective diameter $D_{\rm eff} = (1+2 a)/(2 q)$, which is reported in the horizontal axis.
[Lower panes] $\Si_{eq}$ (left) and
$\Si_j$ (right) for
$p=4$ and $z=3$. In the first case, we varied $q$ at fixed $a$, while in the second
we did the inverse.
}
\label{somesigmad1}
\end{figure}

\begin{figure}[t]
\includegraphics[width=.6\textwidth]{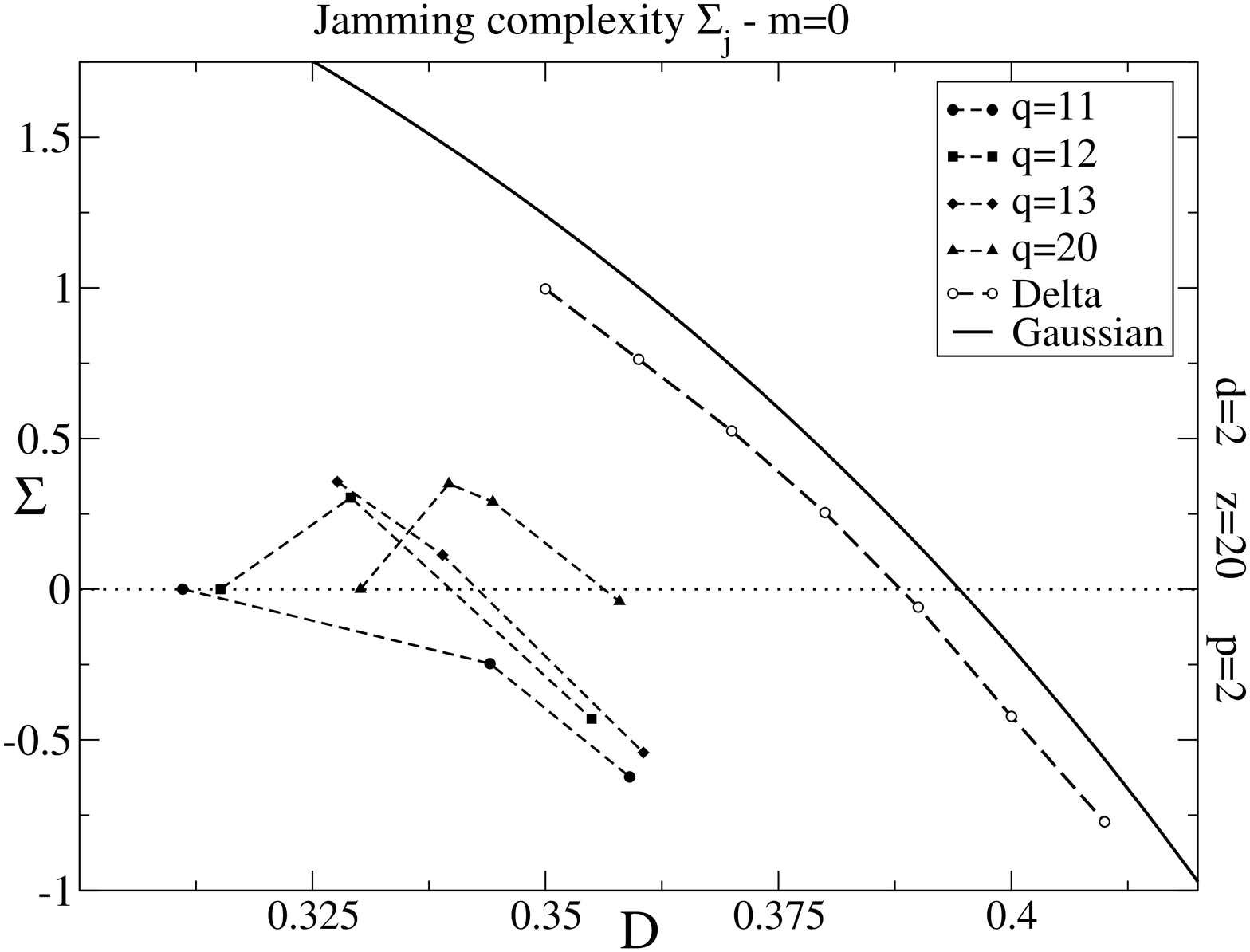}
\caption{
The complexity at $d=2$, $p=2$ and $z=20$, computed with the numerical solution of the
population dynamics algorithm with varying resolution of the
discretization process, is compared to the Gaussian and the Delta approximations. Here we can only use moderate values of $q$,
and because of the geometry of the discretization the effective diameter of the sphere, 
given by Eq.~(\ref{effDd2}) and reported on the horizontal axis,
cannot be varied smoothly. For instance, at $q = 11$ we could not find
a point at positive complexity.
}
\label{somesigmad2}
\end{figure}

\section{Comparison between numerical results and the approximations}
\label{sec:results}

In this section we report the results obtained from the direct numerical calculation with 
discretized space and we compare them with the delta and Gaussian approximations.

\subsection{Complexity}

In Fig.~\ref{somesigmad1} we report the complexities $\Si_{eq}$
 (the complexity at $m=1$  equal to $(1/N)$ time the logarithm of the
 typical 
 number of glass states when configurations are samples uniformly) and $\Si_j$ (the complexity at $m=0$ equal to $(1/N)$ time the logarithm of the
 total number of jammed states)
for several representative cases at $d=1$ where the transition is
discontinuous.
Generically we observe that the delta approximation performs better at $m=0$, while the
Gaussian approximation is more reliable at $m=1$. Both approximations give an upper bound
to the true complexity and therefore 
give values for $D_{\rm K}$ and $D_{\rm GCP}$ that are above the true ones.
Moreover, both approximations miss the dynamical transition since by construction the fields
are assumed to be localized.

Some results for $d=2$ are reported in Fig.~\ref{somesigmad2}.
Here the scaling for $q\to\io$ becomes very difficult because the numerical solution
is computationally demanding and we cannot go beyond $q=20$ for moderate connectivities.
We could perform a systematic investigation only $p=2$ and $z=20$, which is unfortunately
a case where the transition is continuous and the solution might be unstable towards further 
RSB in the glass phase. In this case, at $m=1$ we correctly find a continuous transition at a value
of $D$ which is compatible with the result found from the stability analysis of 
section~\ref{sec:RSstability}.
At $m=0$, we find good agreement with the result of the Gaussian and delta approximation.
Note however that also at $m=0$ the results could be unstable towards further RSB.

\subsection{Phase diagram}

\begin{figure}[t]
\includegraphics[width=.9\textwidth]{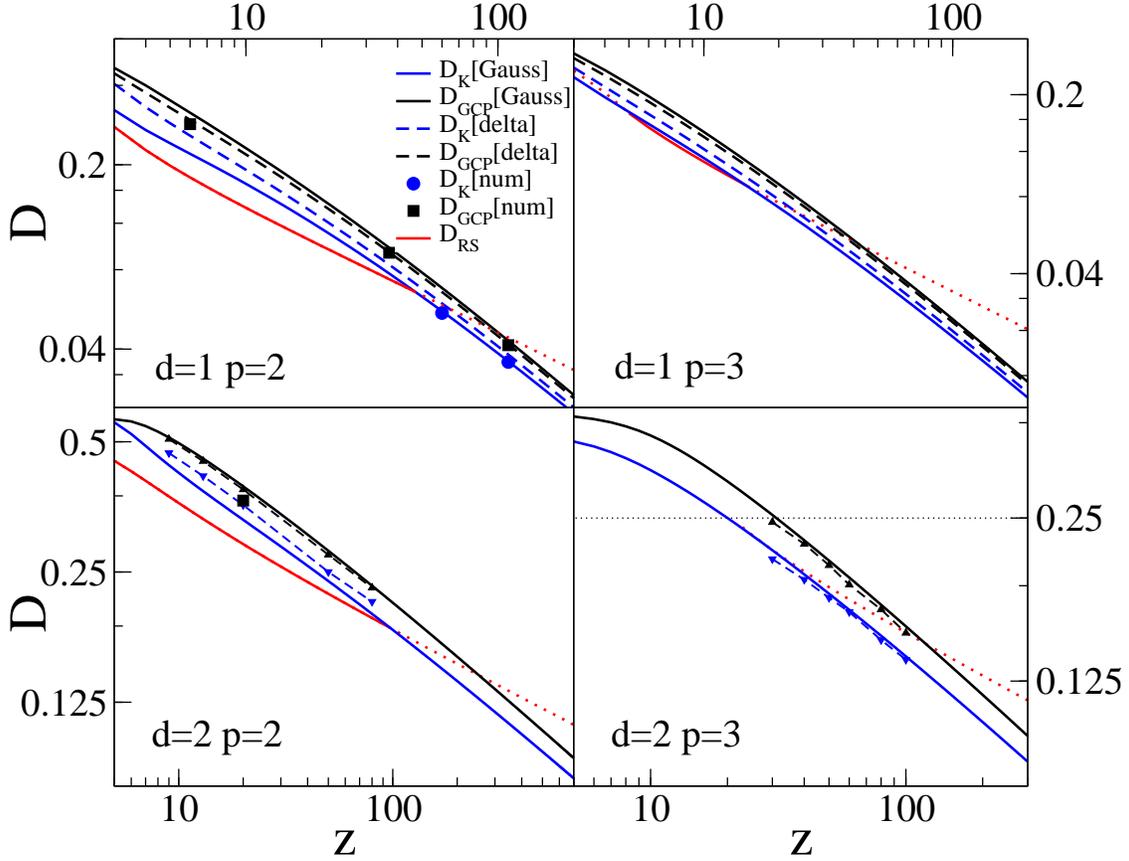}
\caption{
Phase diagrams for $p=2,3$ and $d=1,2$. We compare the results of the Gaussian
and delta approximations with the numerical results obtained directly from a
discretization of the cavity equations. In the lower right panel, the horizontal 
line indicates the value $D=1/4$ above which the calculation of $Z^0_3$ is not valid,
see Eq.~(\ref{Z03}).
}
\label{fig:phasediagrams}
\end{figure}

In Fig.~\ref{fig:phasediagrams} we compare the transition lines obtained by the Gaussian
and delta approximations with the numerical results,
where available. We computed $D_{\rm K}$ and $D_{\rm GCP}$ by performing an extrapolation to $q\to\io$
(which is simple since the corrections are found to be proportional to $1/q$) in some
representative cases where the transition is continuous or discontinuous; the results
are reported in Fig.~\ref{fig:phasediagrams}.
We observe that indeed the Gaussian and delta approximation give consistent
results, which are also consistent with the exact numerical solution and provide upper bounds
to the latter.

Whenever the RS instability $D_{\rm RS} < D_{\rm K}$, the transition is continuous.
This happens generically for small $z$. On increasing $z$, the lines $D_{\rm RS}$ and $D_{\rm K}$
cross and the transition becomes discountinuos. The value $z^*$ where this crossover happens
depends weakly on the space dimension, but it depends strongly on $p$. Indeed we have
$z^* \sim 100$ for $p=2$, while $z^* \sim 20$ for $p=3$ and (as we can infer from Fig.~\ref{gauss})
the transition is always discontinuous for $p>3$.

\section{Correlation function} \label{corr-func}

\subsection{Definition}

As explained in section~\ref{sec:cavityeq}, in the glass phase the cavity equations have
multiple solutions, each describing a different glass state.
Within each state $\a$ we can define a correlation function $g_\a(x,y)$ as follows.
For each box we have:
\beq\begin{split}
g^{(\a)}_a(x,y) & = \frac{1}{p(p-1)} \la \sum_{i\neq j}^{1,p} \d(x-x_i) \d(y-x_j) \ra_{a,\a} \\
& = \frac{1}{p (p-1)} 
\frac{
\int dx^a_1 \cdots dx^a_p \, \psi^{(\a)}_{a,1}(x^a_1) \cdots \psi^{(\a)}_{a,p}(x^a_p) 
\, \chi(x_1^a,\cdots,x_p^a)  \sum_{i\neq j}^{1,p} \d(x-x^a_i) \d(y-x^a_j) }{
\int dx^a_1 \cdots dx^a_p \, \psi^{(\a)}_{a,1}(x^a_1) \cdots \psi^{(\a)}_{a,p}(x^a_p) 
\, \chi(x_1^a,\cdots,x_p^a) }
\ , 
\end{split}\eeq
since the fields $\psi^{(\a)}_{a,i}(x^a_i)$ describe the distribution of the variables adjacents
to box $a$ in absence of the box itself. We now average this quantity over the boxes and over the
states $\a$ with the weight $Z_\a^{m}$. We get
\beq \label{eq:gr} \begin{split}
& g(x,y) = \frac{p}{N z} \sum_{a=1}^{Nz/p} \frac{1}{\sum_\a Z_\a^m} \sum_\a g^{(\a)}_a(x,y) Z_\a^m \\
& = e^{-S_{box}} \int d\PP[\psi_1]\cdots d\PP[\psi_p] \, Z_{box}[\psi_1\cdots \psi_p]^{m-1} \,
\psi_1(x)\psi_2(y) \int \left( \prod_{j=3}^p \psi_{j}(x_j) dx_j \right) \chi(x,y,x_3,\cdots,x_p) 
\end{split}\eeq
Note that in the RS case the above expression reduces to $g^0_p(x,y)$.

We expect that at $m=0$ (close packing),
$g(x,y)$ develops a peak in $|x-y|=D$ describing contacts~\cite{SLN06,DTS05}.
The number of contacts is
\beq
\z = (p-1) \int_{peak} g(0,y) dy \ .
\eeq
The delta peak is also accompanied, in three dimensional sphere packings, by
a square root divergence, $g(r) \sim (r-D)^{-0.5}$~\cite{SLN06,DTS05}, which
we want to investigate here.

Note that in the delta approximation we just get 
\beq
g(x,y) = \frac1{Z^0_p}\int dX_3 \cdots dX_p \chi(x,y,X_3,\cdots,X_p) = g^0_p(x,y)
\eeq
therefore all the structure of the correlation in the packings is lost
in this approximation.

One can show, following \cite{PZ10}, that in the Gaussian approximation, as $A\sim m$ for
$m\to 0$, one gets a delta peak at $r=D$ in the jamming limit, with all particles being non-rattlers
and $\z = 2d$. Therefore this approximation is able to capture some of the peculiar structure
of the correlation.
On the other hand, the square root singularity is missed by the Gaussian 
approximation~\cite{PZ10}. 

Unfortunately, it is very difficult to study the contact peak in the numerical solution
of the cavity equation, because the discretization makes it hard to define a proper
notion of contacts and separate the delta peak contribution from the background.
Therefore in the following we focus on the square root singularity which is also a non-trivial
and somehow unexpected feature of pair correlations at jamming~\cite{SLN06,DTS05}.

\begin{figure}[t]
\includegraphics[width=.9\textwidth]{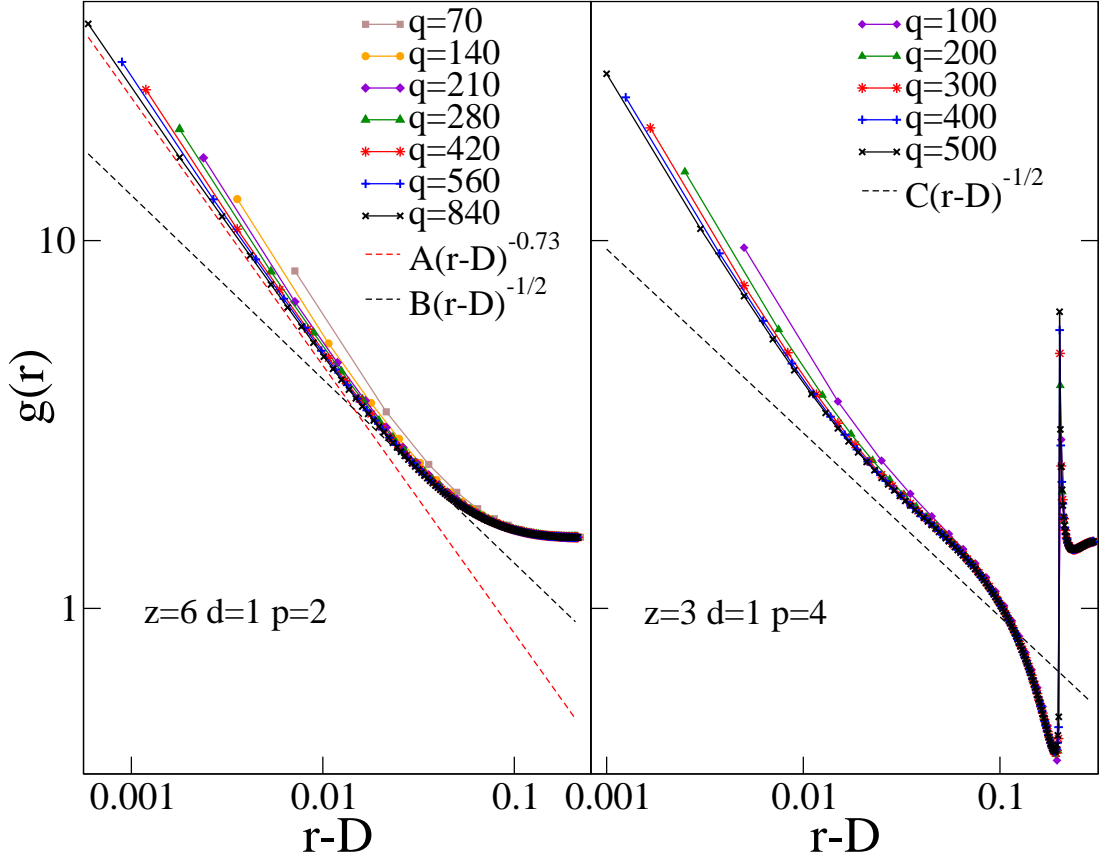}
\caption{
Pair correlation function $g(r)$ at $d=1$, $m=0$ (jamming) and $D \sim D_{\rm GCP}$ (in practice, the closest value to 
$D_{\rm GCP}$ compatible with the discretization). (Left) $p=2$, $z=6$; 
note that in this case the system undergoes a continuous transition and these results might
be unstable towards further RSB. (Right) $p=4$, $z=3$: here the transition is discontinuous.
Note that for $p=4$ we observe an additional
singularity at $r=2D$~\cite{SLN06}.
}
\label{somegr}
\end{figure}

Numerical results are presented in Fig.~\ref{somegr} for the $g(r)$ in one dimension, and two
representative values of $z$ and $p$ where the transition is continuous or discontinuous.
In both cases,
the divergence is compatible with a square root singularity $(r-D)^{-0.5}$ in a range of $r-D$, but at smaller 
$r-D$ the $g(r)$ seems to diverge as $(r-D)^{-\g}$ with an exponent $\g > 0.5$. 
However, in this region the square root divergence is probably mixed
with the contact delta peak, because of the discretization. 
A detailed analysis of this mixing was not possible
because the values of $q$ we could reach were still too small.
Since this investigation is computationally very demanding,
we could not perform a systematic study of the value of the exponent as a function of $p$ and $z$,
nor investigate the more interesting case $d=2$, which is very hard because our discretization does not
preserve the spherical symmetry around the central particle.
We leave a more systematic numerical analysis for future work.

\subsection{Argument for the square-root singularity}

We now present an analytical argument to relate the shape of the cavity fields to the square root
singularity. We focus on $m=0$, and we study the small $r-D$ behavior of $g(r)$ as follows.
We define the quantity
\beq
\Psi(z) =
\int \left( \prod_{j=1}^p \psi_{j}(x_j) dx_j \right) \frac{\chi(x_1,\cdots,x_p)}{\chi(x_1,x_2)} \delta(x_1-x_2-z) \ .
\eeq
Note that $z = x-y \in [-1,1]^d$ but using periodicity one can restrict to
$z\in [-1/2,1/2]^d$ with periodic boundary conditions.
The probability distribution of $\psi$ induces a distribution $\PP[\Psi]$ on $\Psi$.
Then we have
\beq\label{gdiz}
g(z) = \int dx dy g(x,y) \d(x-y-z) =e^{-S_{box}} \int d\PP[\Psi] \frac{\Psi(z) \chi(z)}{\int dz \Psi(z) \chi(z)} 
\theta\left[\int dz \Psi(z) \chi(z)\right]
\ ,
\eeq
where the term $e^{-S_{box}}$ ensures the normalization $\int dz g(z) = 1$.

In the following we restrict for simplicity to $d=1$.
Note that by translational invariance the field $\Psi$ is centered around a random
uniformly distributed position $z_0$, while its shape is encoded by a non-trivial
distribution.
Now assume that with a certain finite probability with respect to the shape distribution,
one has that
\begin{itemize}
\item $\Psi(z)$ vanishes at some finite distance from the center given by
$z_\pm = z_0 \pm \d z_0$. The quantities $z_\pm$ are then also random and uniformly
distributed in $[-1/2,1/2]$;
\item the shape of $\Psi(z)$ around the point
where it vanishes is of the form
\beq\label{shapeexp}
\Psi(z) \sim e^{-\frac{A}{|z-z_\pm|^\a}} \ ;
\eeq
\item and $|z_+-z_-| < 2D$, and $z_+ > D$ (the additional symmetric contribution coming from $z_-$ gives
a factor 2 and will be neglected as all proportionality constants).
\end{itemize}
Then the function $\chi(z) \Psi(z)$
vanishes everywhere except in $[D,z_+]$ where it is given by $\exp\big[-A/(z_+-z)^\a\big]$.
The average over $\PP[\Psi]$, for what concerns this contribution, 
is translated onto an average over $z_+$ and
Eq.~(\ref{gdiz}) becomes :
\beq
g(z) \sim \int d z_+ \frac{e^{-A/(z_+-z)^\a} \th(D\leq z \leq z_+)}
{\int_D^{z_+} dz e^{-A/(z_+-z)^\a}} 
\theta\left[z_+ \geq D\right] = \int_z^C dz_+ \frac{e^{-A/(z_+-z)^\a}}
{\int_D^{z_+} dz e^{-A/(z_+-z)^\a}} \ ,
\eeq
where $C$ is a suitable cutoff that comes from the fact that if $z_+$ is too much larger than
$D$ the approximation Eq.~(\ref{shapeexp}) will break down. We will show that this cutoff does
not matter as the main contribution for $z \to D$ comes from $z_+$ close to $D$.

To simplify notations, we introduce $\l = (z-D)/D$ and $\e = (z_+-D)/D$. Also we define
$a = A/D^\a$ and $c= (C-D)/D$. With these notations we get
\beq
g(\l) \propto \int_\l^c d\e \frac{e^{-a/(\e-\l)^\a}}
{\int_0^{\e} d\l \, e^{-a/(\e-\l)^\a}} \ .
\eeq
The integral in the denominator is dominated by the small $\l$ behavior, that gives
\beq
\int_0^{\e} d\l \, e^{-a/(\e-\l)^\a} \sim \int_0^{\e} d\l \, e^{-\frac{a}{\e^\a} \left(1+\a\frac{\l}{\e}\right)}
=e^{-\frac{a}{\e^\a}} \frac{\e^{\a+1}}{a \a} \ ,
\eeq
and
\beq
g(\l) \propto \int_\l^c d\e \, \e^{-(\a+1)} e^{a \left(\frac{1}{\e^\a}-\frac{1}{(\e-\l)^\a}\right)} \ .
\eeq
We want now to evaluate the integral by a saddle point for $\l \to 0$. We assume (and will check self-consistently)
that the saddle point value $\e^* \gg \l$. Then we can expand for $\l/\e \ll 1$ and
\beq
g(\l) \propto \int_\l^c d\e \, e^{-(\a +1) \log \e - a \a \l \e^{-(\a+1)}} \ .
\eeq
The maximum of the above expression is found at $\e^* = (a \a \l)^{1/(\a+1)} \gg \l$ for small $\l$ as
initially assumed. Substituting this in the expression above one obtains $g(\l) \propto 1/\l$.
To get the correct result we need to compute also the quadratic corrections around the saddle
point. Including these, we finally obtain
\beq
g(\l) \propto \l^{-\frac{\a}{1+\a}} \propto (r-D)^{-\frac{\a}{1+\a}} \ ,
\eeq
\ie a power-law divergence for $z\to D$ with exponent $\in [0,1]$, which is consistent with
the observed exponents in Fig.~\ref{somegr}. 
Note that a square root singularity is obtained for $\a=1$, namely a simple exponential
singularity of the cavity fields.
We checked on our numerical results that indeed
the form of the fields is compatible with the Ansatz (\ref{shapeexp}).

Note that this same argument can be carried out at finite $m$, but in this case we get that 
$g(\l)$ is independent of $\l$ for small $\l$. A more complete analysis should show that at
finite $m$, $g(\l)$ is a power law for $\l \gg O(m^\n)$ with some exponent $\n$, and it crosses
over to a finite value for $\l \ll O(m^\n)$.

\section{Discussion on finite dimensional hard spheres}

One way to recover the normal
hard sphere model from our model is to set $p=2$ and $z=N-1$. However, 
this limit cannot be investigated within the cavity formalism
which is based on taking first the limit $N\to\io$ at finite $z$.
Here the limits $N\to\io$ and $z\to\io$ do not commute, and if
we first send $N\to\io$ and then $z\to\io$ we do not recover the
hard sphere models (a similar behavior is found for the
Bethe lattice spin glass~\cite{cavity}).

Therefore we want here to find a suitable limit that we can take
{\it after} $N\to\io$ to recover the hard sphere model.
As we discussed in the introduction,
one possibility if to set formally $z=1$
and identify $p$ with the number of particles, therefore taking 
$p \gg 1$. Of course, for $z\leq 2$ and finite $p$ the model
does not have any phase transition (it becomes a one-dimensional
model for $z=2$). Therefore, we have to send $p\to\io$ {\it before}
$z$ becomes smaller than $2$.

As a first check, we note that in this limit the RS entropy
\beq
S^{RS} = \frac{z}p \log Z^0_p \to S_{liq}(\f) \ ,
\eeq
where $S_{liq}(\f)$ is the entropy of $d$-dimensional hard spheres
in the thermodynamic limit at fixed packing fraction $\f$.
Actually, there is a problem
with the latter identification, since $Z^0_p$ does not contain a factor $p!$
which should take into account indistinguishability of the particles. This is
indeed to be expected, since we took a formal limit $z\to 1$, but at any
finite $z>1$ the particles are connected to several boxes which makes them
distinguishable. We therefore recover the finite dimensional result for a system
of distinguishable particles. 

Next, we can look at the stability of the RS solution according to Eq.~(\ref{RSstabcondition}).
To compare with standard hard spheres it is crucial to observe that here the box side is one
while $D$ becomes very small for $p\to\io$, in such a way that the packing fraction
$\f = p V_d(D/2) = p V_d(1/2) D^d$ is finite. For $p\to\io$ first and $z\to 1$ after, 
we have $g^0_p(x) \to g_{liq}(x)$,
however $x$ is expressed in units of the box length. If we introduce as usual the distance
$r$ measured in units of the sphere diameter, $r = x/D$, we have (for $k \neq 0$)
\beq
g^0_p(k) = \int dx e^{ikx} g^0_p(x) = D^d \int dr e^{i k D r} g_{liq}(r) = D^d S(k D) \ ,
\eeq
where $S(k D)$ is the structure factor, and the stability condition becomes
\beq
\sqrt{(p-1)(z-1)} D^d | S(k D) | =\sqrt{(p-1)(z-1)} \frac{\f}{p V_d(1/2)} | S(k D) | \leq 1
\eeq
which is always verified for $p\to\io$ since $\f$ and $S(k D)$ are both of order 1.
This is indeed consistent with our investigations of the model at finite $p$ that showed that
the transition is always discontinuous at $p > 4$. We conclude that one cannot observe
a continuous transition in the normal hard spheres model.
This conclusion is consistent with the ones of Biroli and Bouchaud~\cite{BB09} who showed
that indeed replicated liquid theory in finite dimensions does not allow for a 
continuous RSB transition.

We also note that starting from Eqs.~(\ref{SmGauss}), (\ref{Astar_Gauss})
and taking first $p\to\io$ (with $\f = p V_d(D)/2^d$ and
$(p-1) g_p^0(r)= p g_{liq}(r)$) and then $z\to 1$ we recover Eq.~(74) of \cite{PZ10}, 
which is the starting point of the Gaussian 
small cage replica treatment in finite dimensions,
provided we identify again $\lim_{p \to \io} \frac1p \log Z^0_p = S_{liq}(\f)$,
neglecting the problem with the missing $p!$. 
Apart from this {\it caveat}, this is a nice alternative 
derivation of the approximation of \cite{PZ10}, which
is not based on the replica method.

Finally, one could try to take the same formal limit in 
Eq.~(\ref{Smdelta_sampling}) to obtain an alternative approximate expression for $S(m)$
in finite dimensions. Using the relation $Z^0_{p}/Z^0_{p-1} = \la v \ra$, where $v$
is the void space of $p-1$ particles, we obtain for $z\to 1$ (after $p\to \io$):
\beq
S(m) = \log \frac{ \la v^m \ra}{\la v \ra} + \frac1p \log Z^0_p \ .
\eeq
Note however that the void space $v \propto p$, therefore we must rearrange terms as
\beq
S(m) = \log \frac{ \la (v/p)^m \ra}{\la (v/p) \ra} + m \log p + \frac1p \log ( Z^0_p / p^p ) \ .
\eeq
The term $m \log p$ can be dropped since it gives an additive constant to the internal entropy,
and the resulting expression has a well defined $p\to \io$ limit, assuming here that
$\lim_{p \to \io} \frac1p \log (Z^0_p/p!) = S_{liq}(\f)$ (which is however inconsistent with
the previous discussion, for reasons that we do not understand at present).
This expression can in principle be directly computed, even if it is very hard to sample the 
distribution $P(v)$ of void space because at high density $v=0$ 
for most configurations~\cite{STDTS98}.

\section{Conclusions}

In this paper,
we have studied a mean field hard sphere model introduced in~\cite{MKK08}.
The model is similar to a standard hard sphere model, however each sphere
interacts only with a finite and preassigned number of neighbors. The network
of interactions is given by a random graph, such that the model belongs
to the mean field class and is therefore, in principle, exactly solvable via
the cavity method. We therefore derived the cavity equations for the model
and we presented both analytical approximations to their solution and an ``exact''
numerical solution based on a discretization of the space.

We have shown that the analytical approximations give quite reliable results for the phase
diagram and the complexity. In particular, for large enough $z$ and/or $p$, the transition
belongs to the Random First Order class. Therefore, 
as suggested in~\cite{MKK08}, the model displays an ideal glass (Kauzmann) transition
to a glass phase. Following the glass phase upon increasing pressure, one gets to a point
where the pressure diverges, similarly to standard hard spheres close to the so-called
J-point. Given that the model has an exponential number of metastable states, one obtains
a set of J-points spanning a finite range in density.
Overall, the phenomenology of the model in this regime is very close to the one
expected for finite dimensional hard spheres based on mean field approximations, 
see~\cite{PZ10} and Fig.~\ref{dia_totale}.
We found, in particular, that
the Gaussian approximation is very good for the Kauzmann transition 
but tends to overestimate the close packing. This is consistent with what happens
for three-dimensional hard spheres where the Gaussian approximation gives
$\f_{\rm K}\sim 0.62$, which is consistent with numerical estimates,
and $\f_{\rm GCP} \sim 0.68$, while numerical simulations suggest a somewhat smaller
value~\cite{PZ10}.
On the contrary, the delta approximation is very good for close packing 
but tends to overestimate the Kauzmann point. We proposed a formula for the complexity
that is based on the delta approximation and can be computed numerically for three-dimensional
hard spheres. It would be very interesting to do this computation and compare the result
with the Gaussian approximation in that case.

We also found a somehow unexpected result,
that the transition is continuous at small $z$ and $p$. In particular,
for the values of $p=2$ and $z=100$ that have been used in~\cite{MKK08}, the transition
should be very weakly first order.
The physics in presence of a second order transition could be very different.
For instance, in the case of the Sherrington-Kirkpatrick model, the intensive
ground state energy can be found easily: this would correspond to a unique
J-point density. However, the details of this depend on the model, and in particular
on the shape of the complexity function, so we cannot give any conclusive statement.
It would be interesting to investigate better this point by repeating the numerical simulations
of~\cite{MKK08} both in a region where the transition should be strongly second order
(e.g. at $p=2$ and small $z$) and in a region where it should be strongly ``random first order'' (e.g.
for $p=4$ and small $z$).

Finally, we partially investigated the structure of the configurations at jamming.
We computed the correlation function of the model and showed that it
displays a power-law singularity close to contact, at least for $d=1$.
We also gave an analytical argument to explain the mathematical origin of the singularity.
Extending this study
to higher dimension could give insight in the physics that is responsible for this divergence
and hopefully connect it to isostaticity and the presence of soft modes in the spectrum,
as suggested in~\cite{Wyart,WNW05}. Additional numerical simulations could be extremely
useful also in this respect.

{\bf Acknowledgements:} We warmly thank J.~Kurchan,
F.~Krzakala, R.~Mari, G.~Semerjian,
and L.~Zdeborova for many useful and stimulating discussions.

\end{document}